\documentclass[aps,prd,amsfonts,amssymb, amsmath, showpacs, showkeys, a4paper, nofootinbib, superscriptaddress, 11pt, raggedbottom]{revtex4-2}

\usepackage{subcaption,caption,graphicx}

\usepackage{braket, mathtools}
\usepackage{hyperref}
\usepackage{mathrsfs}

\usepackage{color}
\usepackage{dsfont}
\allowdisplaybreaks[1]

\usepackage{tcolorbox}
\definecolor{airforceblue}{rgb}{0.36, 0.54, 0.66}
\definecolor{azure}{rgb}{0.0, 0.5, 1.0}
\newtcolorbox{tdbox}{colback=airforceblue!40!white,colframe=azure!90!black} 
\newcommand{\td}[1]{
	\if\notesOn1
	\begin{tdbox}
		#1
	\end{tdbox}
	\fi
}

\def\notesOn{1}

\newcommand{\sdd}{\mathrm{d}}
\newcommand{\nn}{\nonumber}

\begin{document}

\title{Explicit finite-time illustration of improper unitary evolution for the Klein--Gordon field in de Sitter space}

\author{William~T.~Emond}
\email{wtemond@gmail.com}
\affiliation{Department of Physics and Astronomy, University of Manchester, Manchester M13 9PL, United Kingdom}

\author{~Christian~K\"{a}ding}
\email{christian.kaeding@tuwien.ac.at}
\affiliation{Atominstitut, Technische Universit\"at Wien, Stadionallee 2, A-1020 Wien, Austria}

\author{Peter~Millington}
\email{peter.millington@manchester.ac.uk}
\affiliation{Department of Physics and Astronomy, University of Manchester, Manchester M13 9PL, United Kingdom}


\begin{abstract}
It is known that quantum field theories in curved spacetime suffer from a number of pathologies, including the inability to relate states on different spatial slices by proper unitary time-evolution operators. In this article, we illustrate this issue by describing the canonical quantisation of a free scalar field in de Sitter space and explicitly demonstrating that the vacuum at a given time slice is unitarily inequivalent to that at any other time. In particular, we find that, if both background and Hamiltonian dynamics are taken into account, this inequivalence holds even for infinitesimally small time steps and not only in the asymptotic time limits.
\end{abstract}


\maketitle


\section{Introduction}

The prevailing understanding of our universe and its large-scale evolution is expressed in terms of the $\Lambda$CDM model~\cite{Carroll:2000fy,Peebles:1984ge,Peebles:2002gy}. This, the standard model of cosmology, see Ref.~\cite{Bull:2015stt} for a review, describes a spatially flat universe whose energy budget is composed of baryonic matter, radiation, cold dark matter (CDM) --- modelled as a pressureless perfect fluid --- and a cosmological constant ($\Lambda$). Although there are some tensions (see, e.g., Ref.~\cite{Perivolaropoulos:2021jda}), with the Hubble tension~\cite{Planck:2018vyg, Riess:2019cxk} being a notable example, the predictions of the $\Lambda$CDM model are in excellent agreement with observational data. Given the assumptions of spatial homogeneity and isotropy, the coarse-grained geometry of our universe (on scales $\gtrsim \hspace{-0.25em}100\,{\rm Mpc}$) is described by the Friedmann--Lema\^{i}tre--Robertson--Walker (FLRW) metric~\cite{2012MNRAS.425..116S, Laurent:2016eqo,Ntelis:2017nrj}. Thus, when taken alongside a period of inflation~\cite{Guth:1980zm, Sato:1980yn, Starobinsky:1980te, Linde:1981mu, Albrecht:1982wi, Linde:1983gd}, which precedes the hot big bang phase, both the early- and late-time spacetime geometries of our universe are (quasi-)de Sitter.  In the case of the early-time accelerated expansion, the necessity that inflation ends means that the geometry cannot be pure de Sitter, while, in the case of the late-time accelerated expansion, we might anticipate that the evolution will lead to a geometry that is asymptotically pure de Sitter, at least if dark energy is attributable to a cosmological constant term in the Einstein--Hilbert action of general relativity (GR), as in $\Lambda$CDM.

At low energies or, equivalently, at large distance scales (relative to the Planck scale), gravity can be well described by classical field theory, and GR continues to stand up to experimental and observational scrutiny as the prevailing candidate. Within GR, the cosmological evolution is dictated by the constituents of the Universe's energy budget and their respective equations of state, be they matter, radiation, curvature, or something more exotic. Whereas the quantum description of gravity remains an open question, quantum field theory (QFT) provides us with an elementary description of many of the known and potential constituents of the Universe. Since the scales at which quantum effects become relevant for these constituents can be significantly lower than the Planck scale, we are, from the perspective of effective field theory, justified in considering their quantum dynamics in classical but no-less trivial gravitational backgrounds --- QFT in curved spacetime.

However, there are known issues with the Schr\"odinger-picture description in general curved spacetimes \cite{Helfer:1996my,Torre:1997zs,Torre:1998eq,Cortez:2011sr,Cortez:2012cf,Agullo:2015qqa,Kozhikkal:2023ang}. In particular, it is understood that states on different spatial slices are not related by a unitary transformation, and Ref.~\cite{Giddings:2025abo} extends this issue even to general quantum dynamical geometries. Due to its relevance during the periods of early- and late-time accelerated expansion, QFT in de Sitter space has been a pertinent example. Notably, Ref.~\cite{Anderson:2017hts} has considered a scalar field in a Bunch--Davies vacuum in an expanding spatial section of de Sitter and concluded that there is a constant decay of the vacuum. In the infinite cosmological time limit, this decay leads to a vanishing overlap of two vacuum in- and out-states and consequently to a unitary inequivalence of the vacua. It should be noted that similar problems can also appear in QFTs in flat spacetime. For example, Refs.~\cite{Blasone:1995zc, Blasone:2025atj} demonstrated this in the case of neutrino mixing, where the flavour and mass representations are unitarily inequivalent in the infinite-volume limit. 

In the present article, we follow a similar line of investigation. More precisely, we attempt to canonically quantise the Klein--Gordon field in de Sitter space, to introduce a Fock space and to calculate its transition matrix elements. As expected from the above discussions, this will not go to plan, and the aim of this article is to illustrate through an explicit calculation that the unitary time-evolution operator obtained in this way is not proper in the sense that it maps between unitarily inequivalent representations of the canonical algebra, such that the overlap of vacua at two infinitesimally close time slices is zero in the infinite-volume limit. The immediate consequence is that correlation functions calculated on the so-evolved vacuum states will necessarily violate the Hadamard condition, except at the initial condition \cite{Kay:1988mu}. We interpret our result as a manifestation of Haag's theorem \cite{Haag:1955ev} in de Sitter space QFT.

The remainder of this article is organised as follows. In Sec.~\ref{sec:embedding in dS}, we review the classical Klein--Gordon scalar field theory in de Sitter space. Section~\ref{sec:canonical quantisation} builds upon this to present a canonical quantisation of a scalar field on a classical de Sitter background spacetime. Next, in Sec.~\ref{sec:Fock_basis}, we  construct the Fock basis in the interaction picture and study its time evolution. From computing the vacuum persistence amplitude, we are able to illustrate explicitly that there is zero overlap between vacuum states and infinitesimally close time slices, and that we therefore lack a proper unitary evolution.


\section{Scalar field theory in de Sitter}\label{sec:embedding in dS}

We focus on the following classical action:
\begin{equation}\label{eq: main action}
    S[\phi,g_{\mu\nu}] \ = \ \frac{M^2_{\text{Pl}}}{2}\int\sdd^4 x\,\sqrt{-g}\;\Big(R \: - \: 2\Lambda\Big) \: + \: S_{\phi}[\phi,g_{\mu\nu}] \;,
\end{equation}
with
\begin{equation}\label{eq:scalar action}
	S_\phi[\phi,g_{\mu\nu}] \ = \ \int\sdd^4 x\,\sqrt{-g}\,\mathscr{L}_\phi \ = \ \int\sdd^4 x\,\mathcal{L}_\phi\;,
\end{equation}
and 
\begin{equation}\label{eq:Klein Gordon Lagrangian}
    \mathcal{L}_\phi \ \coloneqq \ \sqrt{-g}\,\mathscr{L}_\phi \ = \ -\sqrt{-g}\,\Big[\frac{1}{2}\,g^{\mu\nu}\nabla_\mu\phi\nabla_\nu\phi \: + \: \frac{1}{2}\xi R \phi^2 \: + \: V(\phi)\Big] \;,
\end{equation}
which describes a real classical scalar field $\phi =\phi(x)$ coupled to Einstein gravity through the metric $g_{\mu\nu}(x)$ and non-minimally to the Ricci scalar $R$, via the dimensionless coupling $\xi$.  Herein, $\Lambda>0$ is the cosmological constant and $V(\phi)$ is the potential of the scalar field. We use the `mostly plus' signature convention for the metric $g_{\mu\nu}$, i.e., $(-,+,+,+)$, and hereafter work in units where the reduced Planck mass $M_{\text{Pl}}=(16\pi G)^{-1/2}$ is set to unity (such that $\phi$ is measured in units of $M_{\text{Pl}}$). 

With a view to quantising the scalar theory, we assume the spacetime $(\mathcal{M},\,g_{\mu\nu})$ is globally hyperbolic, such that $\mathcal{M}$ can be foliated into spacelike Cauchy surfaces. This enables us to introduce the classical momentum field 
\begin{equation}\label{eq:classical canon mom1}
    \pi_\phi \ = \ \frac{\delta S_\phi}{\delta(\nabla_0\phi)} \ = \ \sqrt{|h|}\,n^\mu\nabla_\mu\phi \;,
\end{equation}
conjugate to the scalar field $\phi$, wherein $h_{ij}$ is the induced metric on a spacelike hypersurface $\Sigma$, and $n^\mu= -\sqrt{-g_{00}}\,g^{\mu 0}$ the (future-directed) unit normal vector to this surface. 

From Eqs.~\eqref{eq: main action} and~\eqref{eq:Klein Gordon Lagrangian}, it follows that the metric and scalar-field  equations of motion are 
\begin{subequations}\label{eq:field eqs}
    \begin{gather}
         G_{\mu\nu} \ = \ T^{(\phi)}_{\mu\nu} \: - \: \Lambda g_{\mu\nu} \;, \label{eq:EFE} \\ \left(\Box \: - \: \xi R\right)\phi \: - \: V'(\phi) \ = \  0 \:, \label{eq:SFE}
    \end{gather}
\end{subequations}
where \smash{$G_{\mu\nu}\coloneqq R_{\mu\nu}-g_{\mu\nu}R/2$} is the Einstein tensor (in which $R_{\mu\nu}$ is the Ricci tensor), $\Box\coloneqq g^{\mu\nu}\nabla_\mu\nabla_\nu\phi = \frac{1}{\sqrt{-g}}\partial_\mu\left(\sqrt{-g}g^{\mu\nu}\partial_\nu\phi\right)$ is the d'Alembertian operator acting on $\phi$  (written in terms of the covariant derivative $\nabla_\mu$), $V'(\phi)\coloneqq\partial V(\phi)/\partial \phi$, and
\begin{flalign}
	T^{(\phi)}_{\mu\nu} \ = \ \nabla_\mu\phi\nabla_\nu\phi - \frac{1}{2}g_{\mu\nu}\big[g^{\alpha\beta}\nabla_\alpha\phi\nabla_\beta\phi\: + \: 2V(\phi)\big] + \xi\big[G_{\mu\nu} \: - \: \nabla_\mu\nabla_\nu \: + \: g_{\mu\nu}\Box\big]\phi^2
\end{flalign}
is the (metric) energy-momentum tensor of the scalar field.
The Klein--Gordon inner product $\braket{\ ,\ }$ between any two solutions $\phi_1$ and $\phi_2$ to Eq.~\eqref{eq:SFE} is defined on a spacelike hypersurface, as
\begin{equation}
    \langle\phi_1,\;\phi_2\rangle \ = \ \int_\Sigma\sdd^3x \sqrt{|h|}\,n^\mu\left(\phi_1\nabla_\mu\phi_2 - \phi_2\nabla_\mu\phi_1\right) \ = \ \int_\Sigma\sdd^3x \left(\phi_1\pi_{\phi_{2}} - \phi_2\pi_{\phi_{1}}\right) \;,
\end{equation}
which is independent of the choice of hypersurface $\Sigma$.

In the analysis presented in this paper, we will assume that $\braket{T^{(\phi)}_{\mu\nu}}=0$ and neglect the backreaction of the scalar field on the metric, allowing us to fix the classical background geometry to that of de Sitter space, such that $R=12H^2$, where $H^{-1}$ is the characteristic length scale of de Sitter, whose inverse we identify with the Hubble parameter.

With cosmological contexts in mind, we choose a flat slicing, foliating de Sitter space $\mathcal{M}_{\mathrm{dS}}$ into a continuous set of spatially flat hypersurfaces $\Sigma_\tau$, parametrised by the time coordinate $\tau$, i.e., $\mathcal{M}_{\mathrm{dS}}=\underset{\tau\in\mathbb{R}}{\bigcup}\Sigma_\tau$. 
Furthermore, we shall work in a comoving coordinate system $x^\mu =(\tau,\mathbf{x})$, with the time coordinate given by conformal time $\tau$, which is related to the physical or cosmic time $t$ as 
\begin{equation}\label{eq:conf time cos time relation}
    \tau(t) \ = \ -\int_{t}^{+\infty}\frac{\sdd t'}{a(t')} \ = \ -\frac{1}{a_0 H}\,e^{-Ht} \ \in (-\infty,0)\;,
\end{equation}
where $a(t)= a_0e^{Ht}$ is the cosmic scale factor. The scale factor is then given (with a slight abuse of notation) by 
\begin{equation}\label{eq:scale factor}
	a \ = \ a(\tau) \ = \  - \: \frac{1}{H\tau} \;.
\end{equation}

A particularly desirable aspect of choosing comoving coordinates and conformal time is that the conformal flatness of the background solution to Eq.~\eqref{eq:EFE} is manifest, i.e., 
\begin{equation}\label{eq:metric}
    g_{\mu\nu} \ = \ a^{2}\,\eta_{\mu\nu} \;,
\end{equation}
where $\eta_{\mu\nu}$ is the Minkowski metric. Accordingly, the Klein--Gordon equation [Eq.~\eqref{eq:SFE}] takes the form
\begin{flalign}\label{eq:scalar eom}
	 \partial_\tau^2\phi \: - \: \nabla^2\phi \: + \: 2aH\partial_\tau\phi \: + \: a^2\left(12\xi H^2\phi \: + \: V'(\phi) \right) \ = \ 0 \;,
\end{flalign}
where $\nabla^2$ is the Laplacian, and $\partial_\tau\coloneqq\partial/\partial\tau$. By rescaling the field $\phi=a^{-1}\tilde{\phi}$, we can remove the linear time derivative to arrive at a simplified equation in terms of $\tilde{\phi}$:
\begin{equation}
    \partial_\tau^2\tilde{\phi} \: - \: \nabla^2\tilde{\phi} \: + \: 2a^2H(-1+6\xi)\tilde{\phi} \: + \: a^4\frac{\partial}{\partial\tilde{\phi}}\left(V(a^{-1}\tilde{\phi})\right) \ = \ 0 \;.
\end{equation}
Given this choice of foliation, the conjugate scalar field momentum $\pi_\phi$ becomes
\begin{equation}\label{eq:classical canon mom}
    \pi_\phi \ = \  a^2\,\partial_\tau\phi \;.
\end{equation}
The scalar-field Hamiltonian $H^0_\phi$ can then be constructed as the Legendre transform of $\mathcal{L}_\phi$ (cf.~Eq.~\eqref{eq:Klein Gordon Lagrangian}), giving
\begin{flalign}\label{eq:classical hamiltonian}
	H^0_\phi(\tau) \ &= \ \int\sdd^3\mathbf{x}\, \left(\pi_\phi\,\partial_\tau\phi - \mathcal{L}_\phi\right) \nn\\[-1.5em]\nn\\ &= \ \frac{1}{2}\int\sdd^3\mathbf{x}\, \Big[\frac{1}{a^2}\pi^2_\phi + a^2(\nabla\phi)^2 + 12a^4\xi H^2 \phi^2 + 2a^4V(\phi)\Big] \;.
\end{flalign}
From this we can readily confirm that $H^0_\phi(\tau)$ satisfies the expected Hamilton equations, as required:
\begin{subequations}\label{eq:variational derivative hamilton eqs}
    \begin{flalign}
        \frac{\delta H^0_\phi}{\delta\pi_\phi} \ =& \ \frac{1}{a^2}\pi_\phi \ = \ \partial_\tau\phi \;, \\ \nn\\ \frac{\delta H^0_\phi}{\delta\phi} \ =& \ -a^2\Big[\nabla^2\phi - a^2\left(12\xi H^2 \phi + V'(\phi)\right)\Big] \ = \ -a^2\Big[ \partial_\tau^2\phi + 2aH\partial_\tau\phi\Big] \ = \ -\partial_\tau\pi_\phi \;.
    \end{flalign}
\end{subequations}
In this Hamiltonian formulation, we define the Poisson bracket between two functions $A = A[\phi(x),\pi_\phi(x)]$ and $B = B[\phi(x),\pi_\phi(x)]$ of the canonical variables $\phi$ and $\pi_{\phi}$ at equal times as
\begin{flalign}
    \lbrace A,\, B\rbrace \ \coloneqq& \ \lbrace A[\phi(x),\pi_\phi(x)],\, B[\phi(y),\pi_\phi(y)]\rbrace\Big\vert_{x^0=y^0} \nn\\[-1em] \nn\\ =& \ \int\sdd^3\mathbf{z}\left(\frac{\delta A}{\delta\phi(z)}\frac{\delta B}{\delta\pi_\phi(z)} \: - \: \frac{\delta A}{\delta\pi_\phi(z)}\frac{\delta B}{\delta\phi(z)}\right)\Bigg\vert_{x^0=y^0=z^0} \;.
\end{flalign}
It follows that the canonical variables satisfy 
\begin{flalign}\label{eq:canon variable Poisson algebra}
    \lbrace\phi(\tau,\mathbf{x}),\,&\phi(\tau,\mathbf{y})\rbrace \ = \  \lbrace\pi_\phi(\tau,\mathbf{x}),\,\pi_\phi(\tau,\mathbf{y})\rbrace \ = \ 0\;,\nn\\ &\lbrace\phi(\tau,\mathbf{x}),\,\pi_\phi(\tau,\mathbf{y})\rbrace \ = \ \delta^{(3)}(\mathbf{x}-\mathbf{y}) \;,
\end{flalign}
such that Hamilton's equations can be expressed as
\begin{equation}\label{eq:Hamilton eqs}
    \pi_\phi \ = \ -a^2\lbrace H_\phi^0,\,\phi\rbrace\;,\qquad \partial_\tau\pi_\phi \ = \ \lbrace H_\phi^0,\,\pi_\phi\rbrace \;.
\end{equation}
For a free field with potential
\begin{equation}
\label{eq:freepotential}
    V(\phi)\ =\ \frac{1}{2}\,m^2\phi^2\;,
\end{equation}
the solution of Eq.~\eqref{eq:SFE} can be written in terms of mode functions $v_{\mathbf{k}}(\tau)$ that satisfy
\begin{equation}
\label{eq:standard mode function eom}
    \partial^2_\tau v_{\mathbf{k}}(\tau)+2a(\tau)H\partial_\tau v_{\mathbf{k}}(\tau)+(\mathbf{k}^2 + a^2(\tau)(m^2+12\xi H^2))v_{\mathbf{k}}(\tau)=0\;,
\end{equation}
where $\mathbf{k}$ is the \emph{co-moving} momentum. To extract the instantaneous mode frequency, we redefine the mode functions $v_{\mathbf{k}}(\tau)= a^{-1}\tilde{v}_{\mathbf{k}}(\tau)$, leading to a wave equation of the form 
\begin{equation}
    \partial^2_\tau \tilde{v}_{\mathbf{k}}(\tau) + \omega^2_{\mathbf{k}}(\tau)\tilde{v}_{\mathbf{k}}(\tau) \ = \ 0 \;,
\end{equation}
such that
\begin{equation}
    \omega_{\mathbf{k}}(\tau)=\sqrt{\mathbf{k}^2+ a^2(\tau)m^2_{\textrm{eff}}}\;,
\end{equation}
with effective mass $m^2_{\textrm{eff}}=m^2+12\xi H^2 - \frac{\ddot{a}(\tau)}{a^3(\tau)} = m^2+ 2H^2(6\xi-1)$, where we have used that $\partial_\tau H = 0 = \frac{\ddot{a}(\tau)}{a^2(\tau)}-2a(\tau)H^2$. 

The solutions for the redefined mode functions $\tilde{v}_{\mathbf{k}}(\tau)$ are then
\begin{equation}
	\tilde{v}_{\mathbf{k}}(\tau) \ = \ \sqrt{\frac{H\pi}{4}}\,a(\tau)^{-1/2}\,H^{(2)}_\nu(-|\mathbf{k}|\tau)\;,
\end{equation}
where $H^{(2)}_\nu(z)$ is a Hankel function of the second kind, \smash{$\nu = \sqrt{\frac{9}{4} - \frac{m^2+12H^2\xi}{H^2}}$}, and the normalisation is fixed by the requirement that the vacuum state in the asymptotic past ($\tau\rightarrow -\infty$) is Bunch-Davies. Consequently,
\begin{equation}
    v_\mathbf{k}(\tau) \ = \ \sqrt{\frac{H\pi}{4}}\,a(\tau)^{-3/2}\,H^{(2)}_\nu(-|\mathbf{k}|\tau) \;.
\end{equation}

In the next section, we move to an equivalent description in terms of canonical operator algebra, wherein we will recover the same mode functions.


\section{Canonical quantisation in de Sitter}
\label{sec:canonical quantisation}

Our aim is to construct a canonical operator description of the Klein--Gordon field in de Sitter space. The key difference compared to the familiar canonical quantisation in Minkowski spacetime is the explicit time dependence of the Hamiltonian, which results from the background evolution in our flat slicing of de Sitter space. This will allow us to illustrate the unitary inequivalence of representations of the canonical algebra at different times explicitly.


\subsection{A Schr\"{o}dinger-like picture}
\label{ssec:schrodinger pic section}

We begin by constructing a modified Schr\"{o}dinger picture in which the states carry the time dependence due to the Hamiltonian evolution, while the canonical operators, viz.~the field and conjugate-momentum operators, are time independent. Basis states are usually time independent in the Schr\"{o}dinger picture.  However, since the Hamiltonian carries explicit time dependence, the basis states of this modified Schr\"{o}dinger picture are necessarily time dependent.  In this way, we are able to isolate the effect of this explicit time dependence.

We quantise the classical scalar field theory discussed in the previous section in terms of the canonical field operators $\hat{\phi}_S(\tau,\mathbf{x})$ and $\hat{\pi}_{S}(\tau,\mathbf{x})$. These Schr\"{o}dinger-picture operators are time independent, satisfying
\begin{equation}
	\label{eq:schroedingerdef}
	\partial_\tau\hat{\phi}_S(\tau,\mathbf{x})=0\qquad \text{and}\qquad \partial_\tau\hat{\pi}_S(\tau,\mathbf{x})=0\;.
\end{equation}
Following the canonical approach, we map Poisson brackets of classical fields to commutators of their associated operator fields, i.e., \smash{$\lbrace A,\; B\rbrace \mapsto \frac{1}{i}[A,\; B]$} (in natural units), such that Eq.~\eqref{eq:canon variable Poisson algebra} maps to the canonical operator algebra
\begin{subequations}
\label{eq:canonical field algebra}
	\begin{gather}
		\big[\hat{\phi}_S(\tau,\mathbf{x}),\,\hat{\phi}_S(\tau,\mathbf{y})\big] \ = \ 0 \ = \ \big[\hat{\pi}_{S}(\tau,\mathbf{x}),\,\hat{\pi}_{S}(\tau,\mathbf{y})\big] \;,\\ \big[\hat{\phi}_S(\tau,\mathbf{x}),\,\hat{\pi}_{S}(\tau,\mathbf{y})\big] \ = \ i\delta^{(3)}(\mathbf{x} \: - \: \mathbf{y})\,\hat{\mathds{1}}\;.
	\end{gather}
\end{subequations}
We shall assume that the expectation value of the field operator
\begin{equation}
	\phi_{\psi}(\tau,\mathbf{x})=\braket{\psi(\tau)|\hat{\phi}_S(\tau,\mathbf{x})|\psi(\tau)}
\end{equation}
is a solution of the classical Klein--Gordon equation in de Sitter, with initial conditions determined by $\ket{\psi(\tau_0)}$.

From Eq.~\eqref{eq:classical canon mom}, and imposing that $\partial_\tau\hat{\phi}_S(\tau,\mathbf{x})=0$, we find
\begin{equation}
	\braket{\psi(\tau)|\hat{\pi}_S(\tau,\mathbf{x})|\psi(\tau)} = a^2(\tau)\partial_\tau\phi_{\psi}(\tau,\mathbf{x}) = \braket{\psi(\tau)|ia^2(\tau)[\hat{H}_S^0(\tau),\,\hat{\phi}_S(\tau,\mathbf{x})]|\psi(\tau)} \;,
\end{equation}
from which we can infer that
\begin{equation}
\label{eq:conj mom commutator}
	\hat{\pi}_S(\tau,\mathbf{x}) = ia^2(\tau)[\hat{H}_S^0(\tau),\,\hat{\phi}_S(\tau,\mathbf{x})] \;.
\end{equation}
This agrees with the expected result upon quantising the associated classical relation~\eqref{eq:Hamilton eqs}. Furthermore, combining Eq.~\eqref{eq:schroedingerdef} with the above equation, enforces 
\begin{equation}
\label{eq:conj mom condition}
	\frac{2H}{a(\tau)}\hat{\pi}_S(\tau,\mathbf{x}) + i[\partial_\tau\hat{H}_S^0(\tau),\,\hat{\phi}_S(\tau,\mathbf{x})] \ = \ 0 \;.
\end{equation}
Returning to the Klein--Gordon equation, and substituting the scalar field expectation value, using the condition~\eqref{eq:schroedingerdef}, we can infer the operator equation
\begin{align}
	&\frac{i}{a^2(\tau)}\left[\hat{H}^0_S(\tau),\hat{\pi}_S(\tau,\mathbf{x})\right]+i\left[\partial_\tau\hat{H}^0_S(\tau),\hat{\phi}_S(\tau,\mathbf{x})\right] +\frac{2H}{a(\tau)}\hat{\pi}_S(\tau,\mathbf{x})\nn\\& -\big[\nabla^2-a^2(\tau)(m^2+12\xi H^2)\big]\,\hat{\phi}_S(\tau,\mathbf{x})=0~.
\end{align}
Employing Eq.~\eqref{eq:conj mom condition}, we then have
\begin{equation}
\label{eq:residual KG eq}
	\frac{i}{a^2(\tau)}\left[\hat{H}^0_S(\tau),\hat{\pi}_S(\tau,\mathbf{x})\right]-\big[\nabla^2-a^2(\tau)(m^2+12\xi H^2)\big]\,\hat{\phi}_S(\tau,\mathbf{x})=0~.
\end{equation}

We can construct the Hamiltonian operator by canonically quantising its classical counterpart~\eqref{eq:classical hamiltonian}, giving
\begin{equation}\label{eq:quantised hamiltonian}
	\hat{H}^0_S(\tau) \ = \  \frac{1}{2}\int\sdd^3\mathbf{x}\Big[\frac{1}{a^2(\tau)}\hat{\pi}^2_S(\tau,\mathbf{x}) + a^2(\tau)\left((\nabla\hat{\phi}_S)^2(\tau,\mathbf{x}) + a^2(\tau)(m^2+12\xi H^2)\,\hat{\phi}_S^2(\tau,\mathbf{x})\right)\Big] \;,
\end{equation}
from which we find
\begin{equation}\label{eq:Hamiltonian conj momentum commutator}
    \left[\hat{H}^0_S(\tau),\hat{\pi}_S(\tau,\mathbf{x})\right] \ = \ -ia^2(\tau)\big[\nabla^2-a^2(\tau)(m^2+12\xi H^2)\big]\,\hat{\phi}_S(\tau,\mathbf{x}) \;,
\end{equation}
such that Eq.~\eqref{eq:residual KG eq} is automatically satisfied. Note that Eqs.~\eqref{eq:conj mom commutator} and~\eqref{eq:Hamiltonian conj momentum commutator} agree precisely  with the quantised Hamilton equations (cf. Eqs.~\eqref{eq:variational derivative hamilton eqs} and~\eqref{eq:Hamilton eqs}).

Now, we take the field operators to have the following plane-wave decompositions:
\begin{subequations}
\label{eq:field mode decomp schrodinger1}
	\begin{flalign}
		\hat{\phi}_S(\tau,\mathbf{x}) \ =& \ \int_{\mathbf{k}}\,\varphi_{\mathbf{k}}(\tau)\Big[\hat{b}_{\mathbf{k},S}(\tau)\,e^{i\mathbf{k}\cdot\mathbf{x}} \: + \: \hat{b}^\dagger_{\mathbf{k},S}(\tau)\,e^{-i\mathbf{k}\cdot\mathbf{x}}\Big] \;,\\[0.8em] \hat{\pi}_{S}(\tau,\mathbf{x}) \ =& \ \int_{\mathbf{k}}\,\varpi_{\mathbf{k}}(\tau)\,\Big[\hat{b}_{\mathbf{k},S}(\tau)\,e^{i\mathbf{k}\cdot\mathbf{x}} \: - \: \hat{b}^\dagger_{\mathbf{k},S}(\tau)\,e^{-i\mathbf{k}\cdot\mathbf{x}}\Big] \;,
	\end{flalign}
\end{subequations}
wherein we have used the shorthand notation
\begin{equation}
	\int_{\mathbf{k}}\coloneqq \int\!\frac{{\rm d}^3\mathbf{k}}{(2\pi)^3}~.
\end{equation} 
We require that the creation and annihilation operators $\hat{b}_{\mathbf{k},S}^{\dag}(\tau)$ and $\hat{b}_{\mathbf{k},S}(\tau)$ 
satisfy the equal-time commutation relations
\begin{subequations}
\label{eq:creation/annihilation equal-time commutation relations}
	\begin{gather}
		\big[\hat{b}_{\mathbf{k},S}(\tau),\,\hat{b}_{\mathbf{p},S}(\tau)\big] \ = \ 0 \ = \ \big[\hat{b}^\dagger_{\mathbf{k},S}(\tau),\,\hat{b}^\dagger_{\mathbf{p},S}(\tau)\big] \;,\\ \big[\hat{b}_{\mathbf{k},S}(\tau),\,\hat{b}^\dagger_{\mathbf{p},S}(\tau)\big] \ = \ (2\pi)^3\delta^{(3)}(\mathbf{k} - \mathbf{p})\,\hat{\mathds{1}} \;.
	\end{gather}
\end{subequations}
Using these and the field operator canonical commutation relation~\eqref{eq:canonical field algebra}, enforces 
\begin{equation}
    \varpi_{\mathbf{k}}(\tau) \ = \ \frac{-i}{2\varphi_{\mathbf{k}}(\tau)} \;.
\end{equation}
The Hamiltonian operator $\hat{H}^0_S(\tau)$ can then be expressed in terms of the creation and annihilation operators $\hat{b}_{\mathbf{k},S}^{\dag}(\tau)$ and $\hat{b}_{\mathbf{k},S}(\tau)$, as
\begin{flalign}
\label{eq:hamiltonian mode expansion}
	\hat{H}^0_S(\tau) \ =& \ \frac{1}{2}\int_{\mathbf{k}}\bigg[\mathds{h}^+(\tau)\left(\hat{b}^\dagger_{\mathbf{k},S}(\tau)\hat{b}_{\mathbf{k},S}(\tau) + \hat{b}_{\mathbf{k},S}(\tau)\hat{b}^\dagger_{\mathbf{k},S}(\tau)\right) \nn\\ &\qquad + \mathds{h}^-(\tau)\Big(\hat{b}_{\mathbf{k},S}(\tau)\hat{b}_{-\mathbf{k},S}(\tau) + \hat{b}^\dagger_{\mathbf{k},S}(\tau)\hat{b}^\dagger_{-\mathbf{k},S}(\tau)\Big)\bigg]\;,
\end{flalign}
where 
\begin{equation}
    \mathds{h}^{\pm}(\tau) \ = \ a^{2}(\tau)\varphi^2_{\mathbf{k}}(\tau)\left(\mathbf{k}^2+a^2(\tau)(m^2+12\xi H^2)\right) \pm \frac{1}{4a^{2}(\tau)\varphi^2_{\mathbf{k}}(\tau)}\;.
\end{equation}
Requiring the instantaneous Hamiltonian to be diagonalised (i.e., $\mathds{h}^{-}(\tau) \overset{!}{=} 0$) fixes $\varphi_{\mathbf{k}}(\tau)$ to be
\begin{equation}\label{eq:varphi def}
	\varphi_{\mathbf{k}}(\tau) = \frac{1}{a(\tau)\sqrt{2\omega_\mathbf{k}(\tau)}} \;.
\end{equation}
This results in our canonical field variables having plane-wave decompositions
\begin{subequations}\label{eq:field mode decomp schrodinger}
	\begin{flalign}
		\hat{\phi}_S(\tau,\mathbf{x}) \ =& \ \int_{\mathbf{k}}\,\frac{1}{a(\tau)\sqrt{2\omega_\mathbf{k}(\tau)}}\Big[\hat{b}_{\mathbf{k},S}(\tau)\,e^{i\mathbf{k}\cdot\mathbf{x}} \: + \: \hat{b}^\dagger_{\mathbf{k},S}(\tau)\,e^{-i\mathbf{k}\cdot\mathbf{x}}\Big] \;,\\[0.8em] \hat{\pi}_{S}(\tau,\mathbf{x}) \ =& \ \frac{-i}{\sqrt{2}}\int_{\mathbf{k}}\,a(\tau)\sqrt{\omega_\mathbf{k}(\tau)}\Big[\hat{b}_{\mathbf{k},S}(\tau)\,e^{i\mathbf{k}\cdot\mathbf{x}} \: - \: \hat{b}^\dagger_{\mathbf{k},S}(\tau)\,e^{-i\mathbf{k}\cdot\mathbf{x}}\Big] \;,
	\end{flalign}
\end{subequations}
and reduces Eq.~\eqref{eq:hamiltonian mode expansion} to
\begin{equation}\label{eq:diagonalised hamiltonian}
	\hat{H}^0_S(\tau) \ = \ \frac{1}{2}\int_{\mathbf{k}}\omega_\mathbf{k}(\tau)\,\left(\hat{b}^\dagger_{\mathbf{k},S}(\tau)\hat{b}_{\mathbf{k},S}(\tau)+ \hat{b}_{\mathbf{k},S}(\tau)\hat{b}^\dagger_{\mathbf{k},S}(\tau)\right) \;.
\end{equation}
As is standard, we would like to normal order the Hamiltonian operator to remove the vacuum energy contribution. However, in principle, there is an issue that arises due to the explicit time dependence of the background, which, as we describe in detail below, generates a mixing of creation and annihilation operators as they evolve in time.

By requiring that $\ket{0,\tau}$ corresponds to the instantaneous vacuum state at time $\tau$, we enforce the simultaneous conditions 
\begin{equation}\label{eq:normal ordering condition}
    \hat{b}_\mathbf{k}(\tau) \ket{0,\tau} \ = \ 0 \;\qquad\textrm{and}\qquad \braket{0,\tau|\hat{H}^0_S(\tau)|0,\tau} \ = \ 0\;,
\end{equation}
such that our normal-ordering prescription reorders all creation operators $\hat{b}^\dagger_\mathbf{k}(\tau)$ to the \emph{left} of all annihilation operators $\hat{b}_\mathbf{k}(\tau)$ with respect to the instantaneous vacuum at time $\tau$. Note that this is a \emph{time-independent} definition, since $\tau$ was chosen arbitrarily, and all operators normal-ordered in this manner remain form-invariant for all $\tau$. The ordering is only undone if we wish to express a given operator in a basis of creation and annihilation operators defined with respect to the vacuum at a different time $\tau'\neq \tau$. The normal-ordered Hamiltonian is then (suppressing the normal-ordered notation for convenience)
\begin{equation}\label{eq:normal ordered hamiltonian}
	\hat{H}^0_S(\tau) \ = \ \int_{\mathbf{k}}\omega_\mathbf{k}(\tau)\,\hat{b}^\dagger_{\mathbf{k},S}(\tau)\hat{b}_{\mathbf{k},S}(\tau) \;.
\end{equation}


\subsection{Background evolution via Bogoliubov transformations}
\label{ssec:Background evo and Bogoliubov}

All that remains is to determine the evolution of the creation and annihilation operators $\hat{b}_{\mathbf{k},S}^{\dag}(\tau)$ and $\hat{b}_{\mathbf{k},S}(\tau)$. This can be inferred by means of Eqs.~\eqref{eq:schroedingerdef} to~\eqref{eq:field mode decomp schrodinger}, which enforce that
\begin{subequations}\label{eq:creation/annihilation ops eom}
	\begin{flalign}
		\partial_\tau\hat{b}_{\mathbf{k},S}(\tau) \ =& \ -\left(\partial_\tau\mathrm{ln}(\varphi_{\mathbf{k}}(\tau))\right)\hat{b}^\dagger_{-\mathbf{k},S}(\tau) \ = \ (\partial_\tau\theta_{\mathbf{k}}(\tau))\,\hat{b}^\dagger_{-\mathbf{k},S}(\tau) \;, \\ \partial_\tau\hat{b}^\dagger_{\mathbf{k},S}(\tau) \ =& \ -\left(\partial_\tau\mathrm{ln}(\varphi_{\mathbf{k}}(\tau))\right)\hat{b}_{-\mathbf{k},S}(\tau) \ = \ (\partial_\tau\theta_{\mathbf{k}}(\tau))\,\hat{b}_{-\mathbf{k},S}(\tau)\;,
	\end{flalign}
\end{subequations}
where
\begin{equation}\label{eq:theta eom}
	\partial_\tau\theta_{\mathbf{k}}(\tau) \ = \ -\left(\partial_\tau\mathrm{ln}(\varphi_{\mathbf{k}}(\tau))\right) \ = \ \frac{\partial_\tau\omega_{\mathbf{k}}(\tau)}{2\omega_{\mathbf{k}}(\tau)} + a(\tau)H \ = \ \frac{1}{2}\partial_\tau\mathrm{ln}\left(a^2(\tau)\omega_{\mathbf{k}}(\tau)\right)
\end{equation}
and we have used Eq.~\eqref{eq:varphi def}. Integrating Eq.~\eqref{eq:theta eom}, gives
\begin{equation}
\label{eq:Theta}
	\theta_{\mathbf{k}}(\tau) \ = \ \frac{1}{2}\ln\left(\frac{a^2(\tau)}{a^2(\tau_0)}\frac{\omega_{\mathbf{k}}(\tau)}{\omega_{\mathbf{k}}(\tau_0)}\right) \ = \ \frac{1}{2}\ln\left(\frac{\tau_0^2}{\tau^2}\frac{\omega_{\mathbf{k}}(\tau)}{\omega_{\mathbf{k}}(\tau_0)}\right) \;.
\end{equation}
The solutions to Eq.~\eqref{eq:creation/annihilation ops eom} are then 
\begin{subequations}\label{eq:schrodinger creation/annihilation ops}
	\begin{flalign}
		\hat{b}_{\mathbf{k},S}(\tau) \ =& \ c_\mathbf{k}(\tau)\,\hat{b}_{\mathbf{k},S}(\tau_0) + s_\mathbf{k}(\tau)\,\hat{b}^\dagger_{-\mathbf{k},S}(\tau_0) \;,\\ \hat{b}^\dagger_{\mathbf{k},S}(\tau) \ =& \ c_\mathbf{k}(\tau)\,\hat{b}^\dagger_{\mathbf{k},S}(\tau_0) + s_\mathbf{k}(\tau)\,\hat{b}_{-\mathbf{k},S}(\tau_0) \;,
	\end{flalign}
\end{subequations}
where we have defined 
\begin{equation}\label{eq:cosh sinh notation}
    c_\mathbf{k}(\tau) \ \coloneqq \ \cosh\big(\theta_\mathbf{k}(\tau)\big) \;,\qquad 
    s_\mathbf{k}(\tau) \ \coloneqq \ \sinh\big(\theta_\mathbf{k}(\tau)\big) \;.
\end{equation}
Next, the hyperbolic trigonometric identities lead immediately to
\begin{subequations}
    \begin{flalign}
        c^2_\mathbf{k}(\tau) \: - \: s^2_\mathbf{k}(\tau) \ =& \ 1 \;,\\[0.8em] c_\mathbf{k}(\tau) \: + \: s_\mathbf{k}(\tau) \ =& \ \frac{a(\tau)}{a(\tau_0)}\sqrt{\frac{\omega_{\mathbf{k}}(\tau)}{\omega_{\mathbf{k}}(\tau_0)}} \;.
    \end{flalign}
\end{subequations}
Given these relations, the canonical variables $\hat{\phi}_S$ and $\hat{\pi}_S$, can be recast into explicitly time-independent form: 
\begin{subequations}
    \begin{flalign}
        \hat{\phi}_S(\tau,\mathbf{x}) \ =& \ \int_{\mathbf{k}}\frac{1}{a(\tau_0)\sqrt{2\omega_\mathbf{k}(\tau_0)}}\left[\hat{b}_{\mathbf{k},S}(\tau_0)\,e^{i\mathbf{k}\cdot\mathbf{x}} + \hat{b}^\dagger_{\mathbf{k},S}(\tau_0)\,e^{-i\mathbf{k}\cdot\mathbf{x}}\right] \ = \ \hat{\phi}_S(\tau_0,\mathbf{x})\;,\\[1em] \hat{\pi}_S(\tau,\mathbf{x}) \ =& \ -i\int_{\mathbf{k}}\frac{a(\tau_0)\sqrt{\omega_\mathbf{k}(\tau_0)}}{\sqrt{2}}\left[\hat{b}_{\mathbf{k},S}(\tau_0)\,e^{i\mathbf{k}\cdot\mathbf{x}} - \hat{b}^\dagger_{\mathbf{k},S}(\tau_0)\,e^{-i\mathbf{k}\cdot\mathbf{x}}\right] \ = \ \hat{\pi}_S(\tau_0,\mathbf{x})\;,
    \end{flalign}
\end{subequations}
consistent with the requirement that $\partial_\tau\hat{\phi}_S = 0 = \partial_\tau\hat{\pi}_S$.

In fact, it can be shown (as we shall demonstrate below) that the time evolution of the creation and annihilation operators can be written in terms of a unitary operator
\begin{equation}
\label{eq:S ansatz}
	\hat{S}_S(\tau,\tau_0) \ = \ \mathrm{T}\, e^{-i\int_{\tau_0}^{\tau}{\rm d}\tau'\;\hat{F}_S(\tau',\tau_0)} \;,
\end{equation}
with
\begin{subequations}\label{eq:annihilation/creation bogoliubov}
	\begin{flalign}
		\hat{b}_{\mathbf{k},S}(\tau) \ =& \ \hat{S}_S(\tau,\tau_0)\,\hat{b}_{\mathbf{k},S}(\tau_0)\,\hat{S}^{\dagger}_S(\tau,\tau_0) \;, \\[0.8em] \hat{b}^\dagger_{\mathbf{k},S}(\tau) \ =& \ \hat{S}_S(\tau,\tau_0)\,\hat{b}^\dagger_{\mathbf{k},S}(\tau_0)\,\hat{S}^{\dagger}_S(\tau,\tau_0) \;,
	\end{flalign}
\end{subequations}
and where the Hermitian operator $\hat{F}_S(\tau,\tau_0)$ is determined by
\begin{subequations}
    \begin{flalign}
         -i\Big[\hat{F}_S(\tau,\tau_0),\,\hat{b}_{\mathbf{k},S}(\tau)\Big] \ =& \ \left[\partial_\tau\theta_\mathbf{k}(\tau)\right]\,\hat{b}^\dagger_{-\mathbf{k},S}(\tau) \;,\\[0.5em]  -i\Big[\hat{F}_S(\tau,\tau_0),\,\hat{b}^\dagger_{\mathbf{k},S}(\tau)\Big] \ =& \ \left[\partial_\tau\theta_\mathbf{k}(\tau)\right]\,\hat{b}_{-\mathbf{k},S}(\tau) \;.
    \end{flalign}
\end{subequations}
Its explicit form is
\begin{flalign}\label{eq:unitary time evo generator}
	\hat{F}_S(\tau,\tau_0) \ =& \ \frac{i}{2}\int_{\mathbf{k}}\left[\partial_\tau\theta_\mathbf{k}(\tau)\right]\Big[\hat{b}_{\mathbf{k},S}(\tau)\,\hat{b}_{-\mathbf{k},S}(\tau)  \: - \:\hat{b}^\dagger_{\mathbf{k},S}(\tau)\,\hat{b}^\dagger_{-\mathbf{k},S}(\tau)\Big] \nn\\[0.5em]
	 =& \ \frac{i}{2}\int_{\mathbf{k}}\left[\partial_\tau\theta_\mathbf{k}(\tau)\right]\Big[\hat{b}_{\mathbf{k},S}(\tau_0)\,\hat{b}_{-\mathbf{k},S}(\tau_0) \: - \: \hat{b}^\dagger_{\mathbf{k},S}(\tau_0)\,\hat{b}^\dagger_{-\mathbf{k},S}(\tau_0) \Big] \;.
\end{flalign}
To obtain the second line, we have made use of Eq.~\eqref{eq:schrodinger creation/annihilation ops} to rewrite $\hat{F}_S(\tau,\tau_0)$ in terms of the creation and annihilation operators at the initial time $\tau_0$ and from which we see that $\hat{F}_S(\tau,\tau_0)$ carries only the explicit time dependence encoded in $\theta_{\mathbf{k}}(\tau)$. We can therefore compute the time integral in the exponent and hence the time-ordering in Eq.~\eqref{eq:S ansatz} is trivial, i.e.,
\begin{equation}\label{eq:unordered S}
    \hat{S}_S(\tau,\tau_0) \ = \  e^{-i\,\hat{G}_S(\tau,\tau_0)} \;,
\end{equation}
where $\hat{G}_S(\tau,\tau_0) = \int_{\tau_0}^{\tau}{\rm d}\tau'\;\hat{F}_S(\tau',\tau_0)$.

At this point, we pause to consider the form of the operator $\hat{S}_S(\tau,\tau_0)$ in Eq.~\eqref{eq:unordered S}. Since the work of Haag~\cite{Haag:1955ev}, it has been known that operators of the form $\hat{G}_S(\tau,\tau_0)$ are not proper. This is in the sense that, while preserving the the canonical algebra, the operator $\hat{S}_S(\tau,\tau_0)$ transforms any given state in the Hilbert space into another that has vanishing overlap with any other state in the Hilbert space. Indeed, we will see later that this results in the vacuum state at time $\tau$ having vanishing overlap with the vacuum state at time $\tau+\delta\tau$ in the limit of infinite spatial volume, such that our representations of the scalar quantum field theory on de Sitter space are \emph{unitarily inequivalent} on each time slice. We emphasise that this is a consequence of having a time-dependent Hamiltonian (see, e.g., Ref.~\cite{Giddings:2025abo}), and not unique to scalar quantum field theory on de Sitter space.

Returning to the operator algebra, we can now write the \emph{unequal-time} commutation relations of the creation and annihilation operators (with $\tau>\tau'$), which take the forms
\begin{subequations}
	\begin{flalign}
		\big[\hat{b}_{\mathbf{p},S}(\tau),\, \hat{b}_{\mathbf{k},S}(\tau')\big] \ =& \ s_\mathbf{k}(\tau)\,(2\pi)^3\delta^{(3)}(\mathbf{p} + \mathbf{k}) \ = \ \big[\hat{b}^\dagger_{\mathbf{p},S}(\tau'),\, \hat{b}^\dagger_{\mathbf{k},S}(\tau)\big] \;,\\[0.8em] \big[\hat{b}_{\mathbf{p},S}(\tau),\, \hat{b}^\dagger_{\mathbf{k},S}(\tau')\big] \ =& \ c_\mathbf{k}(\tau)\,(2\pi)^3\delta^{(3)}(\mathbf{p} - \mathbf{k}) \ = \ \big[\hat{b}_{\mathbf{p},S}(\tau'),\, \hat{b}^\dagger_{\mathbf{k},S}(\tau)\big] \;,
	\end{flalign}
\end{subequations}
wherein the non-commutativity of pairs of annihilation operators at different times is indicative of the unitary inequivalence anticipated from the explicit time dependence of the Hamiltonian. Note that these reduce to Eq.~\eqref{eq:creation/annihilation equal-time commutation relations} for equal times. 

We now make use of Ref.~\cite{Mitter:1984gt} to decompose $\hat{S}_S$ [cf.~Eq.~\eqref{eq:unordered S}] into a product of three exponential operators:
\begin{equation}\label{eq:S product identity}
    \hat{S}_S \ = \ e^{-\frac{\textrm{Vol}}{4}\int_{\mathbf{k}}\text{ln}(c_\mathbf{k}^2(\tau))}\,e^{-\frac{1}{2}\int_{\mathbf{k}}t_\mathbf{k}(\tau)\,\hat{K}^+_{\mathbf{k},S}(\tau_0)}\,e^{-\frac{1}{2}\int_{\mathbf{k}}\text{ln}(c_\mathbf{k}^2(\tau))\,\hat{K}^0_{\mathbf{k},S}(\tau_0)}\,e^{\frac{1}{2}\int_{\mathbf{k}}t_\mathbf{k}(\tau)\,\hat{K}^-_{\mathbf{k},S}(\tau_0)} \;,
\end{equation}
where we have defined $t_\mathbf{k}(\tau)\coloneqq\tanh(\theta_\mathbf{k}(\tau))$ and introduced
\begin{subequations}\label{eq:K operators def}
	\begin{flalign}
		\hat{K}^+_{\mathbf{k},S}(\tau) \ \coloneqq& \ \hat{b}^\dagger_{\mathbf{k},S}(\tau)\,\hat{b}^\dagger_{-\mathbf{k},S}(\tau) \;,\\ \hat{K}^-_{\mathbf{k},S}(\tau) \ \coloneqq& \ \hat{b}_{\mathbf{k},S}(\tau)\,\hat{b}_{-\mathbf{k},S}(\tau) \;,\\ \hat{K}^0_{\mathbf{k},S}(\tau) \ \coloneqq& \ \hat{b}^\dagger_{\mathbf{k},S}(\tau)\,\hat{b}_{\mathbf{k},S}(\tau) \;.
	\end{flalign}
\end{subequations}
These operators satisfy the closed algebra
\begin{subequations}\label{eq:K operator algebra}
	\begin{flalign}
		\big[\hat{K}^-_{\mathbf{p},S}(\tau),\,\hat{K}^+_{\mathbf{k},S}(\tau)\big] \ =& \ + 4\,(2\pi)^3\delta^{(3)}(\mathbf{p}-\mathbf{k})\,\hat{K}^0_{\mathbf{k},S}(\tau) \: + \: 2\textrm{Vol}\,(2\pi)^3\delta^{(3)}(\mathbf{p}-\mathbf{k}) \;,\\[0.8em] \big[\hat{K}^0_{\mathbf{p},S}(\tau),\,\hat{K}^{\pm}_{\mathbf{k},S}(\tau)\big] \ =& \ \pm 2\,(2\pi)^3\delta^{(3)}(\mathbf{p}-\mathbf{k})\,\hat{K}^{\pm}_{\mathbf{k},S}(\tau) \;.
	\end{flalign}
\end{subequations}
Note that \smash{$\textrm{Vol} \coloneqq \int_\mathbf{k} e^{i\mathbf{k}\cdot(\mathbf{x}-\mathbf{y})}\big\vert_{\mathbf{x}\rightarrow\mathbf{y}}$} corresponds to the spatial volume within the patch we are considering. As this patch becomes infinitely large, the corresponding volume $\textrm{Vol}\rightarrow\infty$. It is this term which indicates the unitary inequivalence of the representations of the canonical algebra at different times, and we return to this point in Sec.~\ref{sec:Fock_basis}.


\subsection{Time evolution of operators and states}

At this point, it is helpful to clarify what we mean by implicit and explicit time dependence. The time dependence of the background induces an explicit time dependence in the mode frequency $\omega_\mathbf{k}(\tau)$, causing the field normalisation~\eqref{eq:varphi def} to become explicitly time dependent. However, in this construction, the requirement that the canonical variables are time independent further gives rise to an implicit time dependence (that is, generated by some operator $\hat{G}_S$) in the basis operators, viz., the creation and annihilation operators. A generic operator in the theory will therefore be a function of the spacetime coordinates $x^\mu=(\tau,\mathbf{x})$, and the canonical variables $\hat{\phi}_S$ and $\hat{\pi}_S$, thus containing both implicit and explicit time dependence:
\begin{equation}
    \hat{\mathcal{O}}_S(x) \ \coloneqq \ \hat{S}_S(\tau,\tau_0)\hat{\mathcal{O}}_S(x,\hat{\phi}_S(\tau_0,\mathbf{x}),\hat{\pi}_S(\tau_0,\mathbf{x}))\hat{S}_S^\dagger(\tau,\tau_0) \;,
\end{equation}
where $\hat{\mathcal{O}}_S(x,\hat{\phi}_S(\tau_0,\mathbf{x}),\hat{\pi}_S(\tau_0,\mathbf{x}))$ contains the explicit time dependence of the operator. This will generically satisfy an evolution equation of the form 
\begin{equation}\label{eq:generic operator time evo in schrodinger pic}
    \partial_\tau\hat{\mathcal{O}}_S(x) \ = \ -i\Big[\hat{G}_S(\tau,\tau_0),\,\hat{\mathcal{O}}_S(x)\Big] \: + \: \hat{S}_S(\tau,\tau_0)\,(\partial_\tau\hat{\mathcal{O}}_S(x,\hat{\phi}_S(\tau_0,\mathbf{x}),\hat{\pi}_S(\tau_0,\mathbf{x}))\,\hat{S}_S^\dagger(\tau,\tau_0) \;.
\end{equation}
Since state vectors undergo unitary evolution with respect to the Hamiltonian in this construction, the density operator $\hat{\rho}_S(\tau)$ will evolve likewise. Specifically, it will satisfy the Liouville--von Neumann equation
\begin{equation}
	\frac{\sdd}{\sdd\tau}\hat{\rho}_S(\tau) \ = \ -i\big[\hat{H}^0_S(\tau),\,\hat{\rho}_S(\tau)\big] \;,
\end{equation}
with solution
\begin{equation}
	\hat{\rho}_S(\tau) \ = \ \hat{U}(\tau,\tau_0)\hat{\rho}_S(\tau_0)\hat{U}^\dagger(\tau,\tau_0) \;.
\end{equation}
The unitary time-evolution operator has the form
\begin{equation}\label{eq:time-ordered evolution operator}
\hat{U}_0(\tau,\tau_0) \ = \ \mathrm{T}\exp\Big[-i\int_{\tau_0}^{\tau}\sdd\tau'\,\hat{H}^0_S(\tau')\Big] \;,
\end{equation}
where the time ordering is necessitated by the explicit time dependence of the Hamiltonian. The latter means that the Hamiltonian does not commute with itself at different times, i.e., [cf. Eq.~\eqref{eq:normal ordered hamiltonian}]
\begin{flalign}
    \big[\hat{H}^0_S(\tau),\,\hat{H}^0_S(\tau')\big] \ = & \ \int_\mathbf{k}\omega_{\mathbf{k}}(\tau)\,\omega_{\mathbf{k}}(\tau')\Big[\hat{K}^0_{\mathbf{k},S}(\tau),\,\hat{K}^0_{\mathbf{k},S}(\tau')\Big] \nn\\[0.5em] =& \ \int_\mathbf{k}\omega_{\mathbf{k}}(\tau)\,\omega_{\mathbf{k}}(\tau')\sinh\left(2\theta_\mathbf{k}(\tau,\tau')\right)\Big[\hat{K}^-_{\mathbf{k},S}(\tau_0) \: - \: \hat{K}^+_{\mathbf{k},S}(\tau_0)\Big] \ \neq \ 0~.
\end{flalign}
However, given Eqs.~\eqref{eq:schrodinger creation/annihilation ops} and \eqref{eq:K operators def}, we can recast the Hamiltonian in terms of the initial time annihilation and creation operators as
\begin{flalign}\label{eq:Hamiltonian initial operator form}
    \hat{H}^0_S(\tau) \ =& \ \int_\mathbf{k}\omega_{\mathbf{k}}(\tau)\hat{K}^0_{\mathbf{k},S}(\tau) \nn\\[0.5em] =& \ \int_\mathbf{k}\omega_{\mathbf{k}}(\tau)\Big[\mathrm{Vol}\,s_\mathbf{k}^2(\tau) + \left(c_\mathbf{k}^2(\tau)+s_\mathbf{k}^2(\tau)\right)\hat{K}^0_{\mathbf{k},S}(\tau_0)  + s_\mathbf{k}(\tau)c_\mathbf{k}(\tau)\big(\hat{K}^-_{\mathbf{k},S}(\tau_0) + \hat{K}^+_{\mathbf{k},S}(\tau_0)\big)\Big] \;.
\end{flalign}
Note that the Hamiltonian $\hat{H}_S^0(\tau)$ is no longer diagonalised or normal ordered since it is no longer expressed in an instantaneous basis of creation and annihilation operators at time $\tau$. 

We can now carry out the integration over time in the exponent of the evolution operator~\eqref{eq:time-ordered evolution operator}, since it is simply over the $c$-number (time-dependent) coefficients. We can then remove the time ordering, such that
\begin{flalign}\label{eq:untimeordered evo operator}
    \hat{U}_0(\tau,\tau_0) \ =& \ \exp\Big[-i\int_\mathbf{k}\Big(\mathrm{Vol}\,f_\mathbf{k}(\tau,\tau_0) \: + \: g_\mathbf{k}(\tau,\tau_0)\,\hat{K}^0_{\mathbf{k},S}(\tau_0)   + \: h_\mathbf{k}(\tau,\tau_0)\,\big(\hat{K}^-_{\mathbf{k},S}(\tau_0) \: + \: \hat{K}^+_{\mathbf{k},S}(\tau_0)\big)\Big)\Big]  \;,
\end{flalign}
where 
\begin{subequations}\label{eq: f, g and h defs}
    \begin{flalign}
        f_\mathbf{k}(\tau,\tau_0) \ =& \  \int_{\tau_0}^{\tau}\sdd\tau'\,\omega_\mathbf{k}(\tau')s_\mathbf{k}^2(\tau')\ = \ \frac{1}{2}g_\mathbf{k}(\tau,\tau_0) - \frac{1}{2}\int_{\tau_0}^{\tau}\sdd\tau'\,\omega_\mathbf{k}(\tau') \nn\\ =& \ \frac{1}{2}g_\mathbf{k}(\tau,\tau_0) - \frac{1}{2}\left(\tau\omega_\mathbf{k}(\tau) + \tau_0\omega_\mathbf{k}(\tau_0)\right)  + \frac{m}{4H}\,\textrm{ln}\left(\frac{\frac{m}{H\tau} + \omega_\mathbf{k}(\tau)}{\frac{m}{H\tau} - \omega_\mathbf{k}(\tau)}\right) \nn\\ & \ - \frac{m}{4H}\,\textrm{ln}\left(\frac{\frac{m}{H\tau_0} + \omega_\mathbf{k}(\tau_0)}{\frac{m}{H\tau_0} - \omega_\mathbf{k}(\tau_0)}\right)
        \;, \\[0.8em]
        \label{eq:g}
        g_\mathbf{k}(\tau,\tau_0) \ =& \   \int_{\tau_0}^{\tau}\sdd\tau'\,\omega_\mathbf{k}(\tau')\left(c_\mathbf{k}^2(\tau')+s_\mathbf{k}^2(\tau')\right) \ = \   \int_{\tau_0}^{\tau}\sdd\tau'\,\omega_\mathbf{k}(\tau')\cosh(2\theta_\mathbf{k}(\tau')) \nn\\ =& \ \frac{\left(\tau^4 + 2\tau\tau_0^3 - 3\tau_0^4\right) \omega_\mathbf{k}(\tau_0)}{6\tau\tau_0^2} - \frac{m^2 (\tau-\tau_0)^2 (2\tau+\tau_0)}{6H^2\tau^3\tau_0\,\omega_\mathbf{k}(\tau_0)}\;, \\[0.8em]
        \label{eq:h}
        h_\mathbf{k}(\tau,\tau_0) \ =& \ \int_{\tau_0}^{\tau}\sdd\tau'\,\omega_\mathbf{k}(\tau')s_\mathbf{k}(\tau')c_\mathbf{k}(\tau') \ = \ \frac{1}{2}\int_{\tau_0}^{\tau}\sdd\tau'\,\omega_\mathbf{k}(\tau')\sinh(2\theta_\mathbf{k}(\tau')) \nn\\ =& \ \frac{1}{2}g_\mathbf{k}(\tau,\tau_0) - \frac{(\tau^3-\tau_0^3)\omega_\mathbf{k}(\tau_0)}{6\tau_0^2}
        \;.
    \end{flalign}
\end{subequations}
Note the violation of time-translational invariance, which is expected on general grounds (see Ref.~\cite{Giddings:2025abo}).


\subsection{The interaction picture}

In this section, we define an interaction picture in which the field operators evolve with respect to the free part of the Hamiltonian and the states instead evolve with respect to its interaction part.  Note that the interaction and Heisenberg pictures coincide for the free theory that we treat in this work.

The interaction-picture field operators $\hat{\phi}_I(x)$ and $\hat{\pi}_{I}(x)$ are related to their Schr\"{o}dinger-picture counterparts via the unitary map
\begin{equation}\label{eq:Field operator interaction picture map}
    \hat{\phi}_I(x) \ = \ \hat{U}_0^\dagger(\tau,\tau_0)\,\hat{\phi}_S(x)\,\hat{U}_0(\tau,\tau_0)\;,\qquad \hat{\pi}_I(x) \ = \ \hat{U}_0^\dagger(\tau,\tau_0)\,\hat{\pi}_S(x)\,\hat{U}_0(\tau,\tau_0)\;, 
\end{equation}
in which $\hat{U}_0(\tau,\tau_0)$ is the evolution operator given in Eq.~\eqref{eq:time-ordered evolution operator}, with $\hat{H}^0_S(\tau)$ given by Eq.~\eqref{eq:Hamiltonian initial operator form}. These operators can be decomposed as
\begin{subequations}\label{eq:field mode decomp interaction}
	\begin{flalign}
	\hat{\phi}_I(x) \ \coloneqq& \ \hat{\phi}_I(\tau,\mathbf{x}) \ = \ \int_{\mathbf{k}}\,\frac{1}{a(\tau)\sqrt{2\omega_{\mathbf{k}}(\tau)}}\Big[\hat{b}_{\mathbf{k},I}(\tau)e^{i\mathbf{k}\cdot\mathbf{x}} \: + \: \hat{b}^\dagger_{\mathbf{k},I}(\tau)e^{-i\mathbf{k}\cdot\mathbf{x}}\Big] \;,\\[0.8em] \hat{\pi}_{I}(x) \ \coloneqq& \ \hat{\pi}_{I}(\tau,\mathbf{x}) \ = \ \int_{\mathbf{k}}\,(-i)a(\tau)\sqrt{\frac{\omega_{\mathbf{k}}(\tau)}{2}}\,\Big[\hat{b}_{\mathbf{k},I}(\tau)e^{+i\mathbf{k}\cdot\mathbf{x}} \: - \: \hat{b}^\dagger_{\mathbf{k},I}(\tau)e^{-i\mathbf{k}\cdot\mathbf{x}}\Big] \;,
	\end{flalign}
\end{subequations}
where the interaction-picture creation and annihilation operators are obtained via
\begin{subequations}\label{eq:interaction pic creation and annihilation operators}
    \begin{flalign}
        \hat{b}_{\mathbf{k},I}(\tau) \ =& \ \hat{U}_0^\dagger(\tau,\tau_0)\,\hat{b}_{\mathbf{k},S}(\tau)\,\hat{U}_0(\tau,\tau_0) \;,\\[0.8em]
        \hat{b}^\dagger_{\mathbf{k},I}(\tau) \ =& \ \hat{U}_0^\dagger(\tau,\tau_0)\,\hat{b}^\dagger_{\mathbf{k},S}(\tau)\,\hat{U}_0(\tau,\tau_0) \;.
    \end{flalign}
\end{subequations}
This map preserves the canonical algebra, and thus $\hat{b}_{\mathbf{k},I}(\tau)$ and $\hat{b}^\dagger_{\mathbf{k},I}(\tau)$ satisfy
\begin{subequations}\label{eq:Interaction pic canonical algebra}
	\begin{gather}
		\big[\hat{b}_{\mathbf{k},I}(\tau),\,\hat{b}_{\mathbf{p},I}(\tau)\big] \ = \ 0 \ = \ \big[\hat{b}^\dagger_{\mathbf{k},I}(\tau),\,\hat{b}^\dagger_{\mathbf{p},I}(\tau)\big] \;,\\ \big[\hat{b}_{\mathbf{k},I}(\tau),\,\hat{b}^\dagger_{\mathbf{p},I}(\tau)\big] \ = \ (2\pi)^3\delta^{(3)}(\mathbf{k} - \mathbf{p}) \;.
	\end{gather}
\end{subequations}
In addition, we have 
\begin{equation}\label{eq:free evo operator IP}
	\hat{S}_I(\tau,\tau_0) \ = \ \hat{U}^\dagger_0(\tau,\tau_0)\hat{S}_S(\tau,\tau_0)\hat{U}_0(\tau,\tau_0) \ = \ \mathrm{T}\,e^{-i\int_{\tau_0}^\tau\sdd\tau'\,\hat{F}_I(\tau',\tau_0)} \;,
\end{equation}
with
\begin{equation}
	\hat{F}_I(\tau,\tau_0) \ = \ \frac{i}{2}\int_{\mathbf{k}}(\partial_\tau\theta_\mathbf{k}(\tau))\big[\hat{K}^-_{\mathbf{k},I}(\tau) \: - \: \hat{K}^+_{\mathbf{k},I}(\tau)\big] \;.
\end{equation}

A given interaction-picture operator $\hat{\mathcal{O}}_I(x)$ evolves according to
\begin{equation}\label{eq:Interaction pic Heisenberg eq}
	\partial_\tau\mathcal{\hat{O}}_I(x) \ = \ i\big[\hat{H}^0_I(\tau),\,\mathcal{\hat{O}}_I(x)\big] \: + \: \hat{U}_0^\dagger(\tau,\tau_0)\partial_\tau\mathcal{\hat{O}}_S(x)\hat{U}_0(\tau,\tau_0) \;,
\end{equation}
where $\hat{H}^0_I(\tau)$ is the free theory Hamiltonian in the interaction picture, given by
\begin{equation}
	\hat{H}^0_I(\tau) \ = \ \hat{U}_0^\dagger(\tau,\tau_0)\hat{H}_S(\tau)\hat{U}_0(\tau,\tau_0) \ = \ \int_\mathbf{k}\,\omega_{\mathbf{k}}(\tau)\,\hat{K}^0_{\mathbf{k},I}(\tau)\;.
\end{equation}
It is important to note here that \smash{$\partial_\tau\mathcal{\hat{O}}_S(\tau,\mathbf{x})$} is the time derivative of the Schr\"{o}dinger picture operator, as governed by Eq.~\eqref{eq:generic operator time evo in schrodinger pic}.
By definition, \smash{$\partial_\tau\hat{\phi}_S(\tau,\mathbf{x})=0$} and \smash{$\partial_\tau\hat{\pi}_{S}(\tau,\mathbf{x})=0$}, and the interaction-picture field operators therefore evolve subject to
\begin{subequations}\label{eq:field evolution interaction}
	\begin{flalign}
		\partial_\tau\hat{\phi}_I(x) \ =& \ i\big[\hat{H}^0_I(\tau),\,\hat{\phi}_I(x)\big] \;,\\[0.8em] \partial_\tau\hat{\pi}_{I}(x) \ =& \ i\big[\hat{H}^0_I(\tau),\,\hat{\pi}_{I}(x)\big] \;.
	\end{flalign}
\end{subequations}
Equation~\eqref{eq:Interaction picture creation/annihilation eom} reproduces the field equations
\begin{subequations}\label{eq:canonical field eqs. of motion}
	\begin{flalign}
	\partial_\tau\hat{\phi}_I(x) \ =& \  \frac{1}{a^2(\tau)}\,\hat{\pi}_{I}(x) \;,\\[0.8em] \partial_\tau\hat{\pi}_{I}(x) \ =& \ a^2(\tau)\big[\nabla^2 \: - \: a^2(\tau)m_{\textrm{eff}}^2\big]\hat{\phi}_I(x) \;,
	\end{flalign}
\end{subequations}
in agreement with Eqs.~\eqref{eq:scalar eom} and~\eqref{eq:field evolution interaction}. Finally, the interaction-picture creation and annihilation operators evolve according to 
\begin{subequations}\label{eq:Interaction picture creation/annihilation eom}
	\begin{flalign}
		\partial_\tau\hat{b}_{\mathbf{k},I}(\tau) \ =& \  -i\omega_{\mathbf{k}}(\tau)\hat{b}_{\mathbf{k},I}(\tau) \: + \: 
  \left[\partial_\tau\theta_\mathbf{k}(\tau)\right]\,\hat{b}^\dagger_{-\mathbf{k},I}(\tau) \;, \\[0.8em] \partial_\tau\hat{b}^\dagger_{\mathbf{k},I}(\tau) \ =& \  +i\omega_{\mathbf{k}}(\tau)\hat{b}^\dagger_{\mathbf{k},I}(\tau) \: + \: \left[\partial_\tau\theta_\mathbf{k}(\tau)\right]\,\hat{b}_{-\mathbf{k},I}(\tau) \;,
	\end{flalign}
\end{subequations}
with the plane-wave decomposition of the field operators taking the form given in Eq.~\eqref{eq:field mode decomp interaction}.


\subsection{Time evolution of the interaction picture operators}

Returning to the interaction-picture creation and annihilation operators, we would like to solve their evolution equations~\eqref{eq:Interaction picture creation/annihilation eom} in a closed form. We start by writing the interaction-picture operators $\hat{b}_{\mathbf{k},I}(\tau)$ and $\hat{b}^\dagger_{\mathbf{k},I}(\tau)$ in terms of the evolution operator $\hat{U}_{0}(\tau,\tau_0)$ and the background evolution operator $\hat{S}_S(\tau,\tau_0)$, acting on their initial time counterparts $\hat{b}_{\mathbf{k},I}(\tau_0)\equiv\hat{b}_{\mathbf{k},S}(\tau_0)$ and $\hat{b}^\dagger_{\mathbf{k},I}(\tau_0)\equiv\hat{b}^\dagger_{\mathbf{k},S}(\tau_0)$, as
\begin{subequations}
	\begin{flalign}
		\hat{b}_{\mathbf{k},I}(\tau) \ =& \ \hat{\mathcal{W}}^\dagger(\tau,\tau_0)\,\hat{b}_{\mathbf{k},I}(\tau_0)\,\hat{\mathcal{W}}(\tau,\tau_0) \;,\\[1em] \hat{b}^\dagger_{\mathbf{k},I}(\tau) \ =& \ \hat{\mathcal{W}}^\dagger(\tau,\tau_0)\,\hat{b}^\dagger_{\mathbf{k},I}(\tau_0)\,\hat{\mathcal{W}}(\tau,\tau_0) \;,
	\end{flalign}
\end{subequations}
where we have defined
\begin{equation}
    \hat{\mathcal{W}}(\tau,\tau_0) \ \coloneqq \ \hat{S}^\dagger_S(\tau,\tau_0)\hat{U}_0(\tau,\tau_0) \;.
\end{equation}
To proceed, we make use of the generalised Baker--Campbell--Hausdorff (BCH) formula for (anti-)time-ordered products~\cite{Dickinson:2017gtm} to infer that the operations of the two time evolution operators on the initial-time annihilation and creation operators induce interaction-picture solutions at some later time $\tau>\tau_0$ that are a linear combination of $\hat{b}_{\mathbf{k},I}(\tau_0)$ and $\hat{b}^\dagger_{\mathbf{k},I}(\tau_0)$, with $c$-number coefficients that depend on $\omega_\mathbf{k}(\tau)$ and $\theta_\mathbf{k}(\tau)$. In fact, and in view of Eq.~\eqref{eq:canonical field eqs. of motion}, we can arrive at the following expressions:
\begin{subequations}
\label{eq:creation annihilation general solution}
    \begin{flalign}
		\hat{b}_{\mathbf{k},I}(\tau) \ =& \ A_{\mathbf{k}}(\tau,\tau_0)\,\hat{b}_{\mathbf{k},I}(\tau_0) \: + \: B^\ast_{\mathbf{k}}(\tau,\tau_0)\,\hat{b}^\dagger_{-\mathbf{k},I}(\tau_0)  \;, \\[0.5em]  \hat{b}^\dagger_{\mathbf{k},I}(\tau) \ =& \ A^\ast_{\mathbf{k}}(\tau,\tau_0)\,\hat{b}^\dagger_{\mathbf{k},I}(\tau_0) \: + \: B_{\mathbf{k}}(\tau,\tau_0)\,\hat{b}_{-\mathbf{k},I}(\tau_0) \;,
	\end{flalign}
\end{subequations}
where 
\begin{subequations}\label{eq:A B split}
    \begin{flalign}
        A_{\mathbf{k}}(\tau,\tau_0) \ =& \ e^{\theta_{\mathbf{k}}(\tau)}\,\alpha_\mathbf{k}(\tau,\tau_0) \: + \: e^{-\theta_{\mathbf{k}}(\tau)}\,\beta_\mathbf{k}(\tau,\tau_0) \;, \\ B_{\mathbf{k}}(\tau,\tau_0) \ =& \ e^{\theta_{\mathbf{k}}(\tau)}\,\alpha_\mathbf{k}(\tau,\tau_0) \: - \: e^{-\theta_{\mathbf{k}}(\tau)}\,\beta_\mathbf{k}(\tau,\tau_0) \;,
    \end{flalign}
\end{subequations}
with $\alpha_{\mathbf{k}}(\tau,\tau_0)$ and $\beta_{\mathbf{k}}(\tau,\tau_0)$ being functions to be determined. 

The reason for the particular forms of $A_{\mathbf{k}}$ and $B_{\mathbf{k}}$ becomes clear when we plug them back into the scalar-field mode expansion, i.e.,
\begin{flalign}\label{eq:scalar field with creation/annihilation solutions}
	\hat{\phi}_I(x) \ =& \ \int_{\mathbf{k}}\,\frac{1}{a(\tau)\sqrt{2\omega_{\mathbf{k}}(\tau)}}\bigg\lbrace\Big[A_\mathbf{k}(\tau,\tau_0)\,\hat{b}_{\mathbf{k},I}(\tau_0) \: + \: B^\ast_\mathbf{k}(\tau,\tau_0)\,\hat{b}^\dagger_{-\mathbf{k},I}(\tau_0)\Big]\,e^{i\mathbf{k}\cdot\mathbf{x}} \nn\\[0.4em] &\qquad\qquad\qquad\quad\; + \: \Big[A^\ast_\mathbf{k}(\tau,\tau_0)\,\hat{b}^\dagger_{\mathbf{k},I}(\tau_0) \: + \: B_\mathbf{k}(\tau,\tau_0)\,\hat{b}_{-\mathbf{k},I}(\tau_0)\Big]\,e^{-i\mathbf{k}\cdot\mathbf{x}}\bigg\rbrace \nn\\[1em] =& \ \int_{\mathbf{k}}\,\frac{1}{a(\tau)\sqrt{2\omega_{\mathbf{k}}(\tau)}}\Big[\big(A_\mathbf{k}(\tau,\tau_0) \: + \: B_\mathbf{k}(\tau,\tau_0)\big)\,\hat{b}_{\mathbf{k},I}(\tau_0)\,e^{i\mathbf{k}\cdot\mathbf{x}} \nn\\ &\qquad\qquad\qquad\quad + \:  \big(A^\ast_\mathbf{k}(\tau,\tau_0) \: + \: B^\ast_\mathbf{k}(\tau,\tau_0)\big)\,\,\hat{b}^\dagger_{\mathbf{k},I}(\tau_0)\,e^{-i\mathbf{k}\cdot\mathbf{x}}\Big] \nn\\[1em] =& \ \int_{\mathbf{k}}\,\frac{\sqrt{2}\,e^{\theta_\mathbf{k}(\tau)}}{a(\tau)\sqrt{\omega_{\mathbf{k}}(\tau)}}\Big[\alpha_\mathbf{k}(\tau,\tau_0)\,\hat{b}_{\mathbf{k},I}(\tau_0)\,e^{i\mathbf{k}\cdot\mathbf{x}} \: + \: \alpha^\ast_\mathbf{k}(\tau,\tau_0)\,\hat{b}^\dagger_{\mathbf{k},I}(\tau_0)\,e^{-i\mathbf{k}\cdot\mathbf{x}}\Big].
\end{flalign}	
By making the identifications
\begin{equation}\label{eq:v}
	v_\mathbf{k}(\tau) \ \equiv \ \frac{\sqrt{2}\,e^{\theta_\mathbf{k}(\tau)}}{a(\tau)\sqrt{\omega_{\mathbf{k}}(\tau)}}\,\alpha_\mathbf{k}(\tau,\tau_0) \ = \ \frac{\sqrt{2}}{a(\tau_0)\sqrt{\omega_{\mathbf{k}}(\tau_0)}}\,\alpha_\mathbf{k}(\tau,\tau_0)\;,
\end{equation}
we can write
\begin{equation}
 \hat{\phi}_I(x) \equiv \int_{\mathbf{k}}\,\Big[v_\mathbf{k}(\tau)\,\hat{b}_{\mathbf{k},I}(\tau_0)\,e^{i\mathbf{k}\cdot\mathbf{x}} \: + \: v^\ast_\mathbf{k}(\tau)\,\hat{b}^\dagger_{\mathbf{k},I}(\tau_0)\,e^{-i\mathbf{k}\cdot\mathbf{x}}\Big] \;.
\end{equation}
The $v_\mathbf{k}(\tau)$ are precisely the mode functions found in Sec.~\ref{sec:embedding in dS}, satisfiying the following equation of motion (cf. Eq.~\eqref{eq:canonical field eqs. of motion}):
\begin{equation}
    \partial_\tau^2 v_\mathbf{k}(\tau) + 2a(\tau)H\partial_\tau v_\mathbf{k}(\tau) + \omega_\mathbf{k}^2(\tau)v_\mathbf{k}(\tau) \ = \ 0 \;.
\end{equation}
Hence, it is clear that our operator approach is equivalent to the mode-function analysis.

In addition, it is clear from Eq.~\eqref{eq:A B split} that $A_{\mathbf{k}}(\tau,\tau_0)$ and $B_{\mathbf{k}}(\tau,\tau_0)$ must satisfy the initial conditions\footnote{Note that identical relations hold for the complex conjugates $A^\ast_\mathbf{k}(\tau_0,\tau_0)$ and $B^\ast_\mathbf{k}(\tau_0,\tau_0)$.}
\begin{equation}
\label{eq:alpha beta initial conditions}
    A_{\mathbf{k}}(\tau_0,\tau_0) \ = \ 1\;, \quad B_{\mathbf{k}}(\tau_0,\tau_0) \ = \ 0 \;,
\end{equation}
where we have noted that $\theta_{\mathbf{k}}(\tau_0)=0$. This in turn fixes the forms of $\alpha_{\mathbf{k}}(\tau,\tau_0)$ and $\beta_{\mathbf{k}}(\tau,\tau_0)$ at $\tau=\tau_0$. Furthermore, from the canonical algebra~\eqref{eq:Interaction pic canonical algebra}, we can infer that $\alpha_\mathbf{k}(\tau,\tau_0)$ and $\beta_\mathbf{k}(\tau,\tau_0)$ must satisfy the constraint
\begin{align}
\label{eq:AB relation}
	&\lvert A_{\mathbf{k}}(\tau,\tau_0)\rvert^2 \: - \: \lvert B_{\mathbf{k}}(\tau,\tau_0)\rvert^2 \ = \ 4\,\text{Re}\big(\alpha_\mathbf{k}(\tau,\tau_0)\beta^\ast_\mathbf{k}(\tau,\tau_0)\big) \ = \ 1  \;.
\end{align}
To ascertain the exact forms of $\alpha_{\mathbf{k}}(\tau,\tau_0)$ and $\beta_{\mathbf{k}}(\tau,\tau_0)$, we refer back to Eq.~\eqref{eq:canonical field eqs. of motion}, from which we can infer that $\alpha_{\mathbf{k}}(\tau,\tau_0)$ and $\beta_{\mathbf{k}}(\tau,\tau_0)$ must fulfil the following first-order equations:
\begin{subequations}\label{eq:alpha beta eom}
	\begin{flalign}
		\partial_\tau\alpha_{\mathbf{k}}(\tau,\tau_0) \ =& \ -i\omega_{\mathbf{k}}(\tau)\,e^{-2\theta_{\mathbf{k}}(\tau)}\beta_{\mathbf{k}}(\tau,\tau_0) \;,\\[0.5em] \partial_\tau\beta_{\mathbf{k}}(\tau) \ =& \ -i\omega_{\mathbf{k}}(\tau)\,e^{2\theta_{\mathbf{k}}(\tau)}\alpha_{\mathbf{k}}(\tau,\tau_0) \;.
	\end{flalign}
\end{subequations}
Note that these equations provide the following constraint from Eq.~\eqref{eq:AB relation}:
\begin{equation}
    W\big[\alpha_\mathbf{k}(\tau,\tau_0),\alpha^\ast_\mathbf{k}(\tau,\tau_0)\big] \ = \ \frac{i}{2}\,\omega_{\mathbf{k}}(\tau)\,e^{-2\theta_\mathbf{k}(\tau)} \;,
\end{equation}
where $W\big[\alpha_\mathbf{k}(\tau,\tau_0),\alpha^\ast_\mathbf{k}(\tau,\tau_0)\big] \coloneqq 2i\,\text{Im}\big(\alpha_\mathbf{k}(\tau,\tau_0)\partial_\tau\alpha^\ast_\mathbf{k}(\tau,\tau_0)\big)$ is the Wronskian of $\alpha_\mathbf{k}(\tau,\tau_0)$ and its complex conjugate ${\alpha}^\ast_\mathbf{k}(\tau,\tau_0)$.

The system of equations~\eqref{eq:alpha beta eom} can be decoupled by considering the corresponding second-order differential equation for $\alpha_{\mathbf{k}}(\tau,\tau_0)$, whose solution can be used to determine $\beta_{\mathbf{k}}(\tau,\tau_0)$:
\begin{equation}
\label{eq:alpha second order eq}
    \partial^2_\tau\alpha_{\mathbf{k}}(\tau,\tau_0) \: + \: 2a(\tau)H\,\partial_\tau\alpha_{\mathbf{k}}(\tau,\tau_0) \: + \: \omega^2_\mathbf{k}(\tau)\alpha_{\mathbf{k}}(\tau,\tau_0) \ = \ 0 \;, 
\end{equation}
where we have used $\frac{\partial_\tau\omega_\mathbf{k}}{\omega_\mathbf{k}} - 2\partial_\tau\theta_\mathbf{k} = -2aH$. A key observation is that Eq.~\eqref{eq:alpha second order eq} is identical in form to the differential equation for the mode functions $v_{\mathbf{k}}(\tau)$ in the standard analysis [cf.~Eq.~\eqref{eq:standard mode function eom}]. 

As a consequence, we immediately know that Eq.~\eqref{eq:alpha second order eq}, and hence Eq.~\eqref{eq:alpha beta eom}, is readily solvable. Indeed, we find that the general closed-form solutions for $\alpha_{\mathbf{k}}(\tau,\tau_0)$ and $\beta_{\mathbf{k}}(\tau,\tau_0)$, are\footnote{Note that the minus sign in the argument of the Hankel functions takes into account that $\tau\in(-\infty,0)$, and ensures that $(H^{(1)}_\mu(z))^\ast=H^{(2)}_{\mu^\ast}(z)$ for $z\in\mathbb{R}_{\geq 0}$.} 
\begin{subequations}
	\begin{flalign}
		\alpha_{\mathbf{k}}(\tau,\tau_0) \ =& \ (-\tau)^{3/2}\,\big(c_1(\mathbf{k},\tau_0)\,H^{(1)}_\mu(-|\mathbf{k}|\tau) \: + \: c_2(\mathbf{k},\tau_0)\,H^{(2)}_\mu(-|\mathbf{k}|\tau)\big) \;, \\[1em]  	\beta_{\mathbf{k}}(\tau,\tau_0) \ =& \ \frac{i}{\omega_{\mathbf{k}}(\tau)}\,e^{2\theta_\mathbf{k}(\tau)}\,\partial_\tau\alpha_{\mathbf{k}}(\tau,\tau_0) \;,
	\end{flalign}
\end{subequations}
where $H^{(1)}_\mu(z)$ and $H^{(2)}_\mu(z)$ are Hankel functions of the first and second kind, respectively, $c_1$ and $c_2$ are integration constants (to be fixed by the initial conditions), and $\mu\coloneqq\sqrt{\frac{9}{4}-\frac{m^2}{H^2}}$. 

To fix the form of the integration constants $c_1$ and $c_2$, we note that Eq.~\eqref{eq:AB relation} requires 
\begin{equation}
    |c_1(\mathbf{k},\tau_0)|^2e^{\pi\,\text{Im}(\mu)} \: - \: |c_2(\mathbf{k},\tau_0)|^2e^{-\pi\,\text{Im}(\mu)} \ = \ \frac{\pi}{8\tau_0^2}\,\omega_\mathbf{k}(\tau_0)\;,
\end{equation}
where we have taken into account that $\mu$ is either \emph{purely real} or \emph{purely imaginary}, and in such a case, the following identity for the Wronskian of Hankel functions holds:
\begin{equation}
    W[H^{(1)}_{\mu^\ast}(-|\mathbf{k}|\tau),H^{(2)}_{\mu}(-|\mathbf{k}|\tau)] \ = \ \frac{4i}{\pi \tau}\,e^{\pi\text{Im}(\mu)} \;.
\end{equation}
The solutions for $c_1$ and $c_2$ that satisfy the initial conditions and the Wronskian constraint are then:
\begin{subequations}
    \begin{eqnarray}
        c_1(\mathbf{k},\tau_0)  &=& \ -\frac{i\pi}{16(-\tau_0)^{3/2}}\left(2|\mathbf{k}|\tau_0\,H^{(2)}_{\mu -1}(-|\mathbf{k}|\tau_0) \: + \: (2\mu - 2i\tau_0\,\omega_\mathbf{k}(\tau_0)-3)\,H^{(2)}_{\mu}(-|\mathbf{k}|\tau_0)\right) \;,~~~~~~
        \\
        c_2(\mathbf{k},\tau_0)  &=& \ +\frac{i\pi}{16(-\tau_0)^{3/2}}\left(2|\mathbf{k}|\tau_0\,H^{(1)}_{\mu -1}(-|\mathbf{k}|\tau_0) \: + \: (2\mu - 2i\tau_0\,\omega_\mathbf{k}(\tau_0)-3)\,H^{(1)}_{\mu}(-|\mathbf{k}|\tau_0)\right) \;.~~~~~~
    \end{eqnarray}
\end{subequations}
A nice property of these solutions is that they automatically asymptote to the Bunch--Davies (BD) vacuum in the distant past (i.e., $\tau_0\rightarrow -\infty$ with $\tau$ large but not negatively infinite, such that $\tau - \tau_0$ remains finite). Indeed, we find that 
\begin{equation}
    A_{\mathbf{k}}\;\longrightarrow\; e^{i|\mathbf{k}|(\tau - \tau_0)}\quad\text{and}\quad B_{\mathbf{k}}\;\longrightarrow\; 0 \;,
\end{equation}
as required.

Finally, given the general solutions for $\hat{b}_{\mathbf{k},I}(\tau)$ and $\hat{b}^\dagger_{\mathbf{k},I}(\tau)$ [cf.~Eq.\eqref{eq:creation annihilation general solution}] at some arbitrary time $\tau>\tau_0$, we can construct \emph{unequal} time commutation relations as follows:
\begin{subequations}
	\begin{flalign}
		\big[\hat{b}_{\mathbf{k},I}(\tau),\,\hat{b}_{\mathbf{p},I}(\tau')\big] \ =& \ \big(e^{\theta_{\mathbf{k}}(\tau,\tau_0)}\,\alpha^\ast_\mathbf{k}(\tau,\tau_0) \: - \: e^{-\theta_{\mathbf{k}}(\tau,\tau_0)}\,\beta^\ast_\mathbf{k}(\tau,\tau_0)\big)\,(2\pi)^3\delta^{(3)}(\mathbf{p} \: + \: \mathbf{k})  \;, \\[0.8em] \big[\hat{b}_{\mathbf{k},I}(\tau),\,\hat{b}^\dagger_{\mathbf{p},I}(\tau')\big] \ =& \ \big(e^{\theta_{\mathbf{k}}(\tau,\tau_0)}\,\alpha^\ast_\mathbf{k}(\tau,\tau_0) \: + \: e^{-\theta_{\mathbf{k}}(\tau,\tau_0)}\,\beta^\ast_\mathbf{k}(\tau,\tau_0)\big)\,(2\pi)^3\delta^{(3)}(\mathbf{p} \: - \: \mathbf{k}) \;.
	\end{flalign}
\end{subequations}
Note that Eq.~\eqref{eq:alpha beta initial conditions} implies that $\alpha^\ast_\mathbf{k}(\tau_0,\tau_0)=\beta^\ast_\mathbf{k}(\tau_0,\tau_0)=\frac{1}{2}$, such that these reduce to the standard commutation relations~\eqref{eq:creation/annihilation equal-time commutation relations} at equal times, as required.


\section{Constructing a Fock basis in the interaction picture}
\label{sec:Fock_basis}

Now we are in a position to construct the Fock basis and consider its time evolutions, focussing in particular on the instantaneous vacuum states.

We start with the vacuum state on the initial time slice $\tau=\tau_0$, defined such that
\begin{equation}\label{eq:initial vacuum def}
    \hat{b}_{\mathbf{k},I}(\tau_0)\ket{0,\tau_0} \ = \ 0 \;, \qquad \braket{0,\tau_0|0,\tau_0} \ = \ 1 \;,
\end{equation}
where we have normalised the vacuum to unity. Note that the initial-time vacuum state coincides in all three of the standard pictures (Schr\"{o}dinger, Heisenberg, and interaction). Since we are constructing the Fock basis directly in the interaction picture, we shall omit any subscripts on the states themselves, so as not to over-clutter notation.

Single-particle momentum states are then created by acting on the vacuum with the creation operator, i.e.,
\begin{equation}
    \ket{\mathbf{k},\tau_0} \ = \ \hat{b}^\dagger_{\mathbf{k},I}(\tau_0)\ket{0,\tau_0} \;.
\end{equation}
By virtue of the canonical algebra \eqref{eq:Interaction pic canonical algebra}, the single-particle momentum states are normalised, so that
\begin{equation}
    \braket{\mathbf{p},\tau_0|\mathbf{k},\tau_0} \ = \ (2\pi)^3\delta^{(3)}(\mathbf{p}-\mathbf{k}) \;.
\end{equation}
Continuing in this fashion, we find that an $n$-particle state is given by 
\begin{equation}
    \ket{\mathbf{k}_1,\mathbf{k}_2,\ldots,\mathbf{k}_n,\tau_0} \ = \ \hat{b}^\dagger_{\mathbf{k}_1,I}(\tau_0)\hat{b}^\dagger_{\mathbf{k}_2,I}(\tau_0)\cdots\hat{b}^\dagger_{\mathbf{k}_n,I}(\tau_0)\ket{0,\tau_0} \ = \ \prod_{i=1}^n\hat{b}^\dagger_{\mathbf{k}_i,I}(\tau_0)\ket{0,\tau_0} \;.
\end{equation}
Together, all $n$-particle states span the Fock space, with completeness relation
\begin{flalign}\label{eq:initial time completeness relation}
    \hat{\mathds{1}} \ =& \ |0,\tau_0\rangle\langle 0,\tau_0| \: + \: \int_{\mathbf{k}_1}|\mathbf{k}_1,\tau_0\rangle\langle\mathbf{k}_1,\tau_0| \: + \: \frac{1}{2!}\int_{\mathbf{k}_1,\mathbf{k}_2}|\mathbf{k}_1,\mathbf{k}_2,\tau_0\rangle\langle\mathbf{k}_1,\mathbf{k}_2,\tau_0| \: + \: \cdots\nn\\[0.5em] \ =& \ |0,\tau_0\rangle\langle 0,\tau_0| \: + \: \sum_{n=1}^{\infty}\frac{1}{n!}\int_{\mathbf{k}_1,\mathbf{k}_2,\ldots,\mathbf{k}_n}|\mathbf{k}_1,\mathbf{k}_2,\cdots,\mathbf{k}_n,\tau_0\rangle\langle\mathbf{k}_1,\mathbf{k}_2,\cdots,\mathbf{k}_n,\tau_0| \;,
\end{flalign}
where $\hat{\mathds{1}}$ is the identity operator.

Given that we have a mapping from the Schr\"{o}dinger to the interaction picture [cf.~Eq.~\eqref{eq:interaction pic creation and annihilation operators}], the vacuum state in the interaction picture at some arbitrary later time $\tau>\tau_0$ is related to the initial vacuum as
\begin{flalign}\label{eq:vacuum evo interaction pic}
    \ket{0,\tau} \ = \  \hat{\mathcal{W}}^\dagger(\tau,\tau_0)\ket{0,\tau_0}\;.
\end{flalign}
Accordingly, an $n$-particle momentum state evolves as
\begin{equation}
    \ket{\mathbf{k}_1,\mathbf{k}_2,\ldots,\mathbf{k}_n,\tau} \ = \ \hat{\mathcal{W}}^\dagger(\tau,\tau_0)\ket{\mathbf{k}_1,\mathbf{k}_2,\ldots,\mathbf{k}_n,\tau_0} \;.
\end{equation}
By acting from the left and the right of Eq.~\eqref{eq:initial time completeness relation} with $\hat{\mathcal{W}}^\dagger(\tau,\tau_0)$ and $\hat{\mathcal{W}}(\tau,\tau_0)$, respectively, it immediately follows that the completeness relation holds at some arbitrary later time $\tau>\tau_0$, and therefore constitutes a basis for the Fock space at time $\tau$. They need not, however, be unitarily equivalent.


\subsection{Coherent-state basis}

With the aim of providing an explicit illustration of the unitary inequivalence of the Fock basis on different time slices, we introduce an additional basis, which will prove to be particular convenient for determining the form of the vacuum-to-vacuum persistence amplitude. This basis is composed of coherent states.

A coherent state $\ket{\zeta(\mathbf{k},\tau)}\coloneqq\ket{\zeta_{\mathbf{k}}(\tau)}$ is, by definition, the unique eigenstate of the annihilation operator $\hat{b}_{\mathbf{k}}(\tau)$ associated with the eigenvalue $\zeta(\mathbf{k},\tau)\coloneqq\zeta_{\mathbf{k}}(\tau)$:
\begin{equation}\label{eq:coherent state def}
	\hat{b}_{\mathbf{k}}(\tau)\ket{\zeta_{\mathbf{k}}(\tau)} \ = \ \zeta_{\mathbf{k}}(\tau)\ket{\zeta_{\mathbf{k}}(\tau)} \;.
\end{equation}
Coherent states can be related to the Fock basis (constructed in the previous section) as follows. Using the completeness relation~\eqref{eq:initial time completeness relation} at some time $\tau>\tau_0$, we have 
\begin{flalign}
	\ket{\zeta(\tau)} \ =& \ \braket{0,\tau|\zeta(\tau)}\ket{0,\tau} \: + \: \sum_{n=1}^{\infty}\frac{1}{n!}\int_{\mathbf{k}_1,\ldots,\mathbf{k}_n}\braket{\mathbf{k}_1,\ldots,\mathbf{k}_n,\tau|\zeta(\tau)}\ket{\mathbf{k}_1,\ldots,\mathbf{k}_n,\tau} \nn\\[1em] =& \ \braket{0,\tau|\zeta(\tau)}\left[\hat{\mathds{1}} \: + \: \sum_{n=1}^{\infty}\frac{1}{n!}\int_{\mathbf{k}_1,\ldots,\mathbf{k}_n}\zeta_{\mathbf{k}_1}\cdots\zeta_{\mathbf{k}_n}(\tau)\hat{b}^\dagger_{\mathbf{k}_1}(\tau)\cdots\hat{b}^\dagger_{\mathbf{k}_n}(\tau)\right]\ket{0,\tau} \nn\\[1em] =& \ \braket{0,\tau|\zeta(\tau)}\,e^{\int_{\mathbf{k}}\zeta_{\mathbf{k}}(\tau)\hat{b}^\dagger_{\mathbf{k}}(\tau)}\ket{0,\tau} \;.
\end{flalign}

All that remains is to determine $\braket{0,\tau|\zeta(\tau)}$. This can be done, as before, by using the normalisation condition for the coherent state,
\begin{flalign}
	1 \ =& \ \braket{\zeta(\tau)|\zeta(\tau)} \ = \ |\braket{0,\tau|\zeta(\tau)}|^2\bra{0,\tau}e^{\int_{\mathbf{k}}\zeta^\ast_{\mathbf{k}}(\tau)\hat{b}_{\mathbf{k}}(\tau)}\,e^{\int_{\mathbf{p}}\zeta_{\mathbf{p}}(\tau)\hat{b}^\dagger_{\mathbf{p}}(\tau)}\ket{0,\tau} \nn\\[0.5em] =& \ |\braket{0,\tau|\zeta(\tau)}|^2\,e^{\int_{\mathbf{k}}|\zeta_{\mathbf{k}}(\tau)|^2} \;,
\end{flalign}
and therefore
\begin{equation}\label{eq:vacuum coherent state overlap}
	\braket{0,\tau|\zeta(\tau)} \ = \ e^{-\frac{1}{2}\int_{\mathbf{k}}|\zeta_{\mathbf{k}}(\tau)|^2} \;,
\end{equation}
up to an arbitrary phase. As such, a given coherent state can be generated from the vacuum state of the full Fock space as follows: 
\begin{equation}\label{eq:general coherent state}
 	\ket{\zeta(\tau)} \ = \ \ e^{-\frac{1}{2}\int_{\mathbf{k}} |\zeta_{\mathbf{k}}(\tau)|^2}\,e^{\int_{\mathbf{k}} \zeta_{\mathbf{k}}(\tau)\hat{b}_{\mathbf{k}}^\dagger(\tau)}\ket{0,\tau} \ = \ e^{\int_{\mathbf{k}} (\zeta(\mathbf{k},\tau)\hat{b}_{\mathbf{k}}^\dagger(\tau) - \zeta^\ast(\mathbf{k},\tau)\hat{b}_{\mathbf{k}}(\tau))}\ket{0,\tau}  \;.
\end{equation}

Now, as advertised, we can use coherent states to construct a basis for the Fock space. However, we must note that the coherent states are not orthogonal, since
\begin{equation}
	|\braket{\zeta|\zeta'}|^2 \ = \ e^{-\int_{\mathbf{k}}|\zeta_{\mathbf{k}} - \zeta'_{\mathbf{k}}|^2} \;.
\end{equation}
Instead, they form an overcomplete basis, and any coherent state can be expanded in terms of all other coherent states (using Eq.~\eqref{eq:coherent completeness}) as
\begin{equation}
	\ket{\zeta'} \ = \ 	\int_{\mathbb{C}}\mathcal{D}\zeta^\ast\mathcal{D}\zeta\,\ket{\zeta}\braket{\zeta|\zeta'} \ = \ \int_{\mathbb{C}}\mathcal{D}\zeta^\ast\mathcal{D}\zeta\,e^{-\frac{1}{2}\int_\mathbf{k}\left(|\zeta_\mathbf{k}-\zeta'_\mathbf{k}|^2 + 2i\text{Im}(\zeta_\mathbf{k}\zeta_\mathbf{k}^{\prime\ast})\right)}\ket{\zeta} \;.
\end{equation}
Despite their overcompleteness, the states nevertheless satisfy a completeness relation
\begin{equation}\label{eq:coherent completeness}
	\hat{\mathds{1}} \ = \ \int_{\mathbb{C}}\mathcal{D}\zeta^\ast\mathcal{D}\zeta\,\ket{\zeta}\bra{\zeta} \;,
\end{equation}
where we have introduced the functional integration measure
\begin{equation}
 	\mathcal{D}\zeta^\ast\mathcal{D}\zeta \ \coloneqq \ \prod_\mathbf{k}\frac{\sdd\zeta^\ast_{\mathbf{k}}\,\sdd\zeta_{\mathbf{k}}}{2\pi i} \;.
\end{equation}

The coherent-state basis will provide us with a convenient way of determining expectation values of operators $\hat{\mathcal{O}}$. Indeed, given that it is always possible to cast a product of creation and annihilation operators into the form $\hat{\mathcal{O}}_{\mathbf{k},I}=\sum_{n,m}c_{nm}\hat{b}_{\mathbf{k},I}^n(\hat{b}_{\mathbf{k},I}^\dagger)^m$, the expectation value of this operator with respect to some state $\ket{\psi}$ can be written in terms of coherent states as
\begin{flalign}
	\braket{\psi|\hat{\mathcal{O}}_{\mathbf{k},I}|\psi} \ = \ \sum_{nm}c_{nm}\braket{\psi|\hat{b}_{\mathbf{k},I}^n(\hat{b}_{\mathbf{k},I}^\dagger)^m|\psi} \ =& \ \sum_{nm}c_{nm}\int_{\mathbb{C}}\mathcal{D}\zeta^\ast\mathcal{D}\zeta\,\braket{\psi|\hat{b}_{\mathbf{k},I}^n|\zeta}\braket{\zeta|(\hat{b}_{\mathbf{k},I}^\dagger)^m|\psi} \nn\\[0.5em] =& \ \sum_{nm}c_{nm}\int_{\mathbb{C}}\mathcal{D}\zeta^\ast\mathcal{D}\zeta\,\zeta_{\mathbf{k}}^n(\zeta_{\mathbf{k}}^\ast)^m|\braket{\zeta|\psi}|^2 \;.
\end{flalign}
We now return to the main goal of this section:~the calculation of the vacuum-to-vacuum persistence amplitude for the free scalar field theory.


\subsection{Vacuum persistence amplitude}

Using the technology established above, we can calculate the probability of finding a patch of the Universe in the Bunch--Davies vacuum, at some time $\tau>\tau_0$, given that it was in this state at the initial time $\tau_0$. To do so, we need to compute the relevant amplitude $\braket{0,\tau|0,\tau_0}$, which we shall refer to as the vacuum persistence amplitude (VPA).

Using Eqs.~\eqref{eq:vacuum evo interaction pic} and~\eqref{eq:coherent completeness}, the VPA can be determined explicitly, as
\begin{flalign}\label{eq:VVT amp}
    \braket{0,\tau|0,\tau_0} \ =& \ \bra{0,\tau_0}\hat{\mathcal{W}}(\tau,\tau_0)\ket{0,\tau_0} \ = \ \bra{0,\tau_0}\hat{S}_S^\dagger(\tau,\tau_0)\hat{U}_0(\tau,\tau_0)\ket{0,\tau_0} \nn\\[0.5em] =& \ \int_{\mathbb{C}}\mathcal{D}\zeta^\ast(\tau_0)\mathcal{D}\zeta(\tau_0)\,\bra{0,\tau_0}\hat{S}^\dagger_S(\tau,\tau_0)\ket{\zeta(\tau_0)}\bra{\zeta(\tau_0)}\hat{U}_0(\tau,\tau_0)\ket{0,\tau_0}\;.
\end{flalign}
We focus on the two correlation functions, $\bra{0,\tau_0}\hat{S}^\dagger_S(\tau,\tau_0)\ket{\zeta(\tau_0)}$ and $\bra{\zeta(\tau_0)}\hat{U}_0(\tau,\tau_0)\ket{0,\tau_0}$,  separately.

Starting with the leftmost (and employing the conjugate transpose of the identity~\eqref{eq:S product identity}), we have
\begin{flalign}\label{eq:coherent state vacuum overlap}
    \bra{0,\tau_0}\hat{S}^\dagger_S(\tau,\tau_0)\ket{\zeta(\tau_0)} \ =& \ e^{-\frac{\textrm{Vol}}{4}\int_{\mathbf{k}}\text{ln}(c_\mathbf{k}^2(\tau))}\nn\\ & \times\bra{0,\tau_0}e^{\frac{1}{2}\int_{\mathbf{k}}t_\mathbf{k}(\tau)\,\hat{K}^+_{\mathbf{k},I}(\tau_0)}\,e^{-\frac{1}{2}\int_{\mathbf{k}}\text{ln}(c_\mathbf{k}^2(\tau))\,\hat{K}^0_{\mathbf{k},I}(\tau_0)}\,e^{-\frac{1}{2}\int_{\mathbf{k}}t_\mathbf{k}(\tau)\,\hat{K}^-_{\mathbf{k},I}(\tau_0)}\ket{\zeta(\tau_0)} \nn\\[0.5em] =& \ e^{-\frac{\text{Vol}}{2}\int_{\mathbf{k}}\text{ln}(c_\mathbf{k}(\tau,\tau_0))}\,e^{-\frac{1}{2}\int_{\mathbf{k}}\left(\text{ln}(c^2_\mathbf{k}(\tau))\,|\zeta_{\mathbf{k}}(\tau_0)|^2 \: + \: 2i\,t_\mathbf{k}(\tau)\,\textrm{Im}(\zeta_{\mathbf{k}}^2(\tau_0))\right)}\braket{0,\tau_0|\zeta(\tau_0)} \nn\\[0.5em] =& \ e^{-\frac{\text{Vol}}{2}\int_{\mathbf{k}}\text{ln}(c_\mathbf{k}(\tau))}\,e^{-i\int_{\mathbf{k}}t_\mathbf{k}(\tau)\,\textrm{Im}(\zeta_{\mathbf{k}}^2(\tau_0))}\,e^{-\frac{1}{2}\int_{\mathbf{k}}\left(1 \: + \: \text{ln}(c^2_\mathbf{k}(\tau))\right)|\zeta_{\mathbf{k}}(\tau_0)|^2} \;.
\end{flalign}
Comparing this to Eq.~\eqref{eq:VVT amp} at $\tau=\tau_0$, we see that normalization of the initial vacuum to unity enforces
\begin{equation}
\int_{\mathbb{C}}\mathcal{D}\zeta^\ast(\tau_0)\mathcal{D}\zeta(\tau_0)\,e^{-\int_{\mathbf{k}}|\zeta_{\mathbf{k}}(\tau_0)|^2} \ = \ 1 \;.
\end{equation}

\begin{figure} [htbp]
\centering
   \includegraphics[scale=0.5]{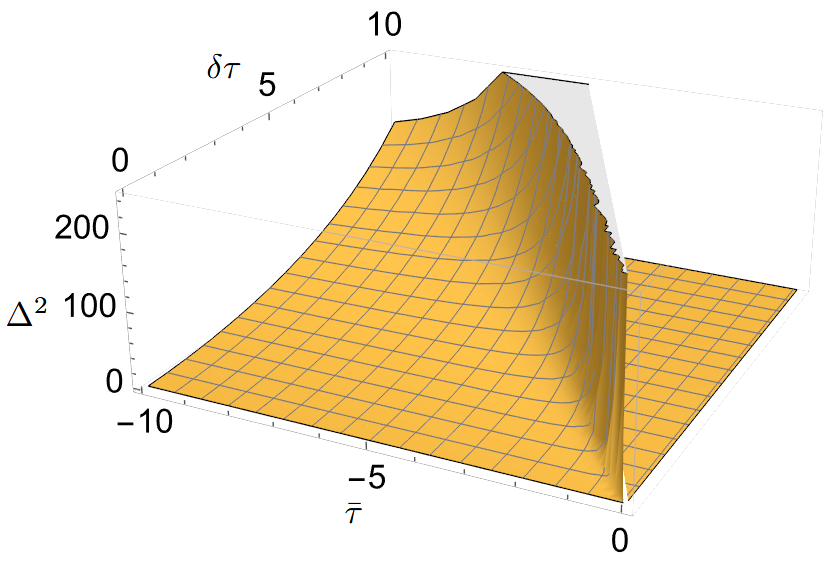}
\caption{Depiction of the function $\Delta^2(\bar{\tau},\delta\tau)$ for $\mathfrak{m} =1$, illustrating that $\Delta^2\geq 0$ for $\tau\geq \tau_0$ ($\bar{\tau}<-\delta\tau/2$).}
    \label{fig:PlotDelta}
\end{figure}

Next, we consider the rightmost term. We first recall that $\hat{U}_0(\tau,\tau_0)$ is of the form
\begin{flalign}
\label{eq:inverse evo op unordered}
    \hat{U}_0(\tau,\tau_0) \ =& \  \exp\Big[-i\int_\mathbf{k}\Big(\mathrm{Vol}\,f_\mathbf{k}(\tau,\tau_0) \: + \: g_\mathbf{k}(\tau,\tau_0)\,\hat{K}^0_{\mathbf{k},I}(\tau_0) \nn\\ &\qquad\qquad\quad + \: h_\mathbf{k}(\tau,\tau_0)\,\big(\hat{K}^-_{\mathbf{k},I}(\tau_0) \: + \: \hat{K}^+_{\mathbf{k},I}(\tau_0)\big)\Big)\Big]\;,
\end{flalign}
where $f_\mathbf{k}(\tau,\tau_0)$, $g_\mathbf{k}(\tau,\tau_0)$ and $h_\mathbf{k}(\tau,\tau_0)$ are given by Eq.~\eqref{eq: f, g and h defs}. Since the operators $\hat{K}^0_{\mathbf{k},I}(\tau_0)$, $\hat{K}^-_{\mathbf{k},I}(\tau_0)$, and $\hat{K}^+_{\mathbf{k},I}(\tau_0)$ form a closed algebra (cf.~Eq.\eqref{eq:K operator algebra}, which is preserved by the unitary mapping to the interaction picture), it is possible to factorise the exponential in Eq.~\eqref{eq:inverse evo op unordered} into the form (see Ref.~\cite{Mitter:1984gt} and Appendix~\ref{sec:factorisation derivation} for further details on the derivation of this factorisation)
\begin{flalign}\label{eq:evo op factorisation}
    \hat{U}_0(\tau,\tau_0) \ =& \ \exp\Big[-i\text{Vol}\int_\mathbf{k}\left(\tilde{\lambda}_\mathbf{k}(\tau,\tau_0) + f_\mathbf{k}(\tau,\tau_0)\right)\Big]\,\exp\Big[-i\int_\mathbf{k}\lambda^{+}_\mathbf{k}(\tau,\tau_0)\,\hat{K}^+_{\mathbf{k},I}(\tau_0)\Big]\nn\\ &\ \times\exp\Big[-i\int_\mathbf{k}\lambda^{0}_\mathbf{k}(\tau,\tau_0)\,\hat{K}^0_{\mathbf{k},I}(\tau_0)\Big]\,\exp\Big[-i\int_\mathbf{k}\lambda^{-}_\mathbf{k}(\tau,\tau_0)\,\hat{K}^-_{\mathbf{k},I}(\tau_0)\Big] \;,
\end{flalign}
where
\begin{subequations}\label{eq:lambda sols}
    \begin{flalign}
        \lambda^{0}_\mathbf{k} \ =& \ i\,\text{ln}\left(\cos\left(\Delta_\mathbf{k}\right) - \frac{i\,g_{\mathbf{k}}}{\Delta_\mathbf{k}}\sin\left(\Delta_\mathbf{k}\right)\right) \;,\\[0.8em]
         \lambda^{\pm}_\mathbf{k} \ =& \   \frac{h_\mathbf{k}\tan(\Delta_\mathbf{k})}{\Delta_\mathbf{k} - ig_\mathbf{k}\tan(\Delta_\mathbf{k})} \;,\\[0.8em]
         \tilde{\lambda}_\mathbf{k} \ =& \ \frac{1}{2}\left(\lambda^0_{\mathbf{k}}-g_\mathbf{k} \right)\;,\\[0.8em]
         \Delta_{\mathbf{k}}^2\ \coloneqq& \ g_{\mathbf{k}}^2-4h_{\mathbf{k}}^2\;.
    \end{flalign}
\end{subequations}
We will confirm later, see Fig.~\ref{fig:PlotDelta}, that $\Delta_{\mathbf{k}}^2\geq 0$ for $\tau\geq\tau_0$.

With Eq.~\eqref{eq:evo op factorisation} in hand, we can straightforwardly determine $\bra{\zeta(\tau_0)}\hat{U}(\tau,\tau_0)\ket{0,\tau_0}$ as
\begin{flalign}
    \bra{\zeta(\tau_0)}\hat{U}_0(\tau,\tau_0)\ket{0,\tau_0} \ =& \ e^{-i\text{Vol}\int_\mathbf{k}\left(\tilde{\lambda}_\mathbf{k}(\tau,\tau_0)+f_\mathbf{k}(\tau,\tau_0)\right)}\,e^{-i\int_\mathbf{k}\lambda^{+}_\mathbf{k}(\tau,\tau_0)\,(\zeta^\ast_{\mathbf{k}}(\tau_0))^2}\braket{\zeta(\tau_0)|0,\tau_0} \nn\\[0.5em] =& \ e^{-i\text{Vol}\int_\mathbf{k}\left(\tilde{\lambda}_\mathbf{k}(\tau,\tau_0)+f_\mathbf{k}(\tau,\tau_0)\right)}\,e^{-i\int_\mathbf{k}\lambda^{+}_\mathbf{k}(\tau,\tau_0)\,(\zeta^\ast_{\mathbf{k}}(\tau_0))^2}e^{-\frac{1}{2}\int_\mathbf{k}|\zeta_\mathbf{k}(\tau_0)|^2}\;.
\end{flalign}
Combining this with the expression in Eq.~\eqref{eq:coherent state vacuum overlap}, and inserting them both into Eq.~\eqref{eq:VVT amp}, we arrive at the result
\begin{flalign}\label{eq:vacuum overlap}
    \braket{0,\tau|0,\tau_0} \ =& \ e^{-\frac{\text{Vol}}{2}\int_{\mathbf{k}}\left(\text{ln}(c_\mathbf{k}(\tau)) \: + \: 2i\tilde{\lambda}_\mathbf{k}(\tau,\tau_0) + 2if_\mathbf{k}(\tau,\tau_0)\right)}Q(\tau,\tau_0) \;,
\end{flalign}
where
\begin{flalign}\label{eq:Q int}
    Q(\tau,\tau_0) \ =& \ \int_{\mathbb{C}}\mathcal{D}\zeta^\ast(\tau_0)\mathcal{D}\zeta(\tau_0)\, e^{-\int_{\mathbf{k}}\left((1+\textrm{ln}(c_\mathbf{k}(\tau)) |\zeta_{\mathbf{k}}(\tau_0)|^2 \: + \: \frac{1}{2}t_\mathbf{k}(\tau)\,\textrm{Im}(\zeta_{\mathbf{k}}^2(\tau_0)) \: + \: \lambda^{+}_\mathbf{k}(\tau,\tau_0)\,(\zeta^\ast_{\mathbf{k}}(\tau_0))^2\right)} \nn\\[0.5em] =& \ \int_{\mathbb{C}}\mathcal{D}\chi^\dagger(\tau_0)\mathcal{D}\chi(\tau_0)\, \textrm{exp}\bigg[-\int_{\mathbf{k}}\mathbf{\chi}^\dagger_\mathbf{k}(\tau_0)\,\mathbb{M}_\mathbf{k}(\tau,\tau_0)\,\mathbf{\chi}_\mathbf{k}(\tau_0)\bigg] \;,
\end{flalign}
with 
\begin{equation}
    \mathbf{\chi}_\mathbf{k}(\tau_0) \ = \ \Bigg(\begin{matrix} \zeta^\ast_{\mathbf{k}}(\tau_0) \\ \zeta_{\mathbf{k}}(\tau_0)\end{matrix}\Bigg)
\end{equation}
and
\begin{equation}
    \mathbb{M}_\mathbf{k}(\tau,\tau_0) \ = \ \begin{pmatrix} \frac{1}{2}\left(1+\mathrm{ln}(c_\mathbf{k}(\tau)\right) && \lambda^{+}_\mathbf{k}(\tau,\tau_0) - \frac{1}{4i}t_\mathbf{k}(\tau) \\ \frac{1}{4i}t_\mathbf{k}(\tau) && \frac{1}{2}\left(1+\mathrm{ln}(c_\mathbf{k}(\tau))\right)\end{pmatrix} \;.
\end{equation}
Consequently, the associated probability density is
\begin{equation}\label{eq:transition probability}
    \lvert\braket{0,\tau|0,\tau_0}\rvert^2 \ = \ e^{-\text{Vol}\int_{\mathbf{k}}\left(\text{ln}(c_\mathbf{k}(\tau)) \: - \: 2\textrm{Im}\tilde{\lambda}_\mathbf{k}(\tau,\tau_0) \right)}\lvert Q(\tau,\tau_0)\rvert^2 \;.
\end{equation}

The full VPA~\eqref{eq:vacuum overlap} is a rather unwieldy quantity to evaluate. However, the result~\eqref{eq:transition probability} is sufficient to determine whether the probability is finite or not. In the infinite-volume limit, there are three scenarios, which are determined by the value of the integral
\begin{equation}
    \label{eq:integral}
    I = \int_{\mathbf{k}}\left(\text{ln}\, c_\mathbf{k}(\tau) \: - \: 2\,{\rm Im}\,\tilde{\lambda}_\mathbf{k}(\tau,\tau_0)\right)\;,
\end{equation}
which multiplies the $\textrm{Vol}$ factor:
\begin{itemize}
    \item If $I=0$, then the overlap is non-zero and finite.
    \item If $I<0$, then the overlap is divergent.
    \item If $I>0$, then the overlap is zero, and we conclude that the Fock bases are unitarily inequivalent at different times.
\end{itemize}

In the limit $\tau_0\to-\infty$ and $\tau\to 0$, i.e., for asymptotic initial and final vacua, ${\rm Im}\,\tilde{\lambda}_{\mathbf{k}}\to 0$, and we recover previous results, see, e.g., Ref.~\cite{Anderson:2017hts}. Returning to finite conformal times, the real part of the integrand is 
\begin{eqnarray}
    \label{eq:LogArg}
         \frac{1}{2}\,\textrm{ln}\left(\frac{c^2_\mathbf{k}}{\cos^2(\Delta_\mathbf{k}) + g^2_\mathbf{k}\,\textrm{sinc}^2(\Delta_\mathbf{k})}\right) \;, \label{eq:vol real part}
\end{eqnarray}
where $\textrm{sinc}(\Delta_\mathbf{k})=\frac{\sin(\Delta_\mathbf{k})}{\Delta_\mathbf{k}}$. Proving that the argument of the logarithm on the right-hand side of Eq.~(\ref{eq:LogArg}) is always greater than or equal to unity appears to be intricate. However, we can provide evidence for the positivity of the right-hand side of Eq.~(\ref{eq:LogArg}) by considering particular benchmark points in the parameter space. To this end, we introduce $m/H =: \mathfrak{m}$, and $\bar{\tau} := |\mathbf{k}|(\tau + \tau_0)/2$ and $\delta\tau := |\mathbf{k}|(\tau - \tau_0)$, such that $\tau = (\bar{\tau} + \delta\tau/2)/|\mathbf{k}|$ and $\tau_0 = (\bar{\tau} - \delta\tau/2)/|\mathbf{k}|$. We also introduce the function
\begin{eqnarray}
\label{eq:aleph}
\aleph &:=& 
c^2_\mathbf{k} - 
\cos^2(\Delta_\mathbf{k}) - g^2_\mathbf{k}\,\textrm{sinc}^2(\Delta_\mathbf{k})
    ~,
\end{eqnarray}
and, after setting $\xi \equiv 1/6$, we have 
\begin{eqnarray}
  \omega_\mathbf{k}(\tau) 
  &=&
  |\mathbf{k}| \sqrt{1 + \frac{\mathfrak{m}^2}{(\bar{\tau} + \delta\tau/2)^2}}
  ~,~~~
  \omega_\mathbf{k}(\tau_0) 
  =
  |\mathbf{k}| \sqrt{1 + \frac{\mathfrak{m}^2}{(\bar{\tau} - \delta\tau/2)^2}}~.
\end{eqnarray}
Using the new parametrization, Eqs.~(\ref{eq:Theta}), (\ref{eq:g}) and (\ref{eq:h}) can be expressed as
\begin{eqnarray}
    \theta_\mathbf{k}(\tau) 
    &=&
    \frac{1}{2}\ln\left(\frac{(\bar{\tau} - \delta\tau/2)^2}{(\bar{\tau} + \delta\tau/2)^2}\frac{\sqrt{1 + \frac{\mathfrak{m}^2}{(\bar{\tau} + \delta\tau/2)^2}}}{\sqrt{1 + \frac{\mathfrak{m}^2}{(\bar{\tau} - \delta\tau/2)^2}}}\right)~,
\end{eqnarray}
\begin{eqnarray}
    g_\mathbf{k}(\tau,\tau_0) 
    &=&
    \frac{\left[(\bar{\tau} + \delta\tau/2)^4 + 2(\bar{\tau} + \delta\tau/2)(\bar{\tau} - \delta\tau/2)^3 - 3(\bar{\tau} - \delta\tau/2)^4\right] }{6(\bar{\tau} + \delta\tau/2)(\bar{\tau} - \delta\tau/2)^2}\sqrt{1 + \frac{\mathfrak{m}^2}{(\bar{\tau} - \delta\tau/2)^2}}
    \nonumber
    \\
    &&
    - 
    \frac{\mathfrak{m}^2 \delta\tau^2 (3\bar{\tau} + \delta\tau/2)}{6(\bar{\tau} + \delta\tau/2)^3(\bar{\tau} - \delta\tau/2)\,\sqrt{1 + \frac{\mathfrak{m}^2}{(\bar{\tau} - \delta\tau/2)^2}}}
\end{eqnarray}
and
\begin{eqnarray}
    h_\mathbf{k}(\tau,\tau_0) 
    &=&   
    \frac{1}{2}g_\mathbf{k}(\tau,\tau_0) - \frac{(\bar{\tau} + \delta\tau/2)^3-(\bar{\tau} - \delta\tau/2)^3}{6(\bar{\tau} - \delta\tau/2)^2}
    \sqrt{1 + \frac{\mathfrak{m}^2}{(\bar{\tau} - \delta\tau/2)^2}}~.
\end{eqnarray}

\begin{figure} [htbp]
\centering
    \subfloat[][]{\includegraphics[scale=0.35]{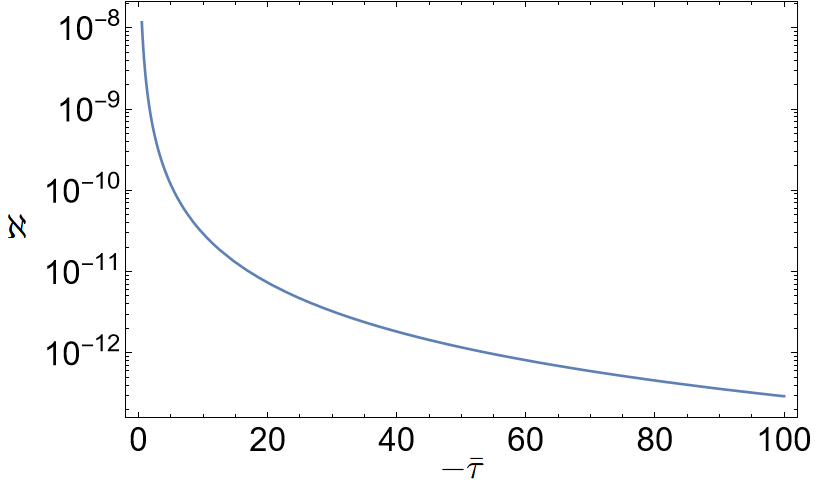}}
    \subfloat[][]{\includegraphics[scale=0.35]{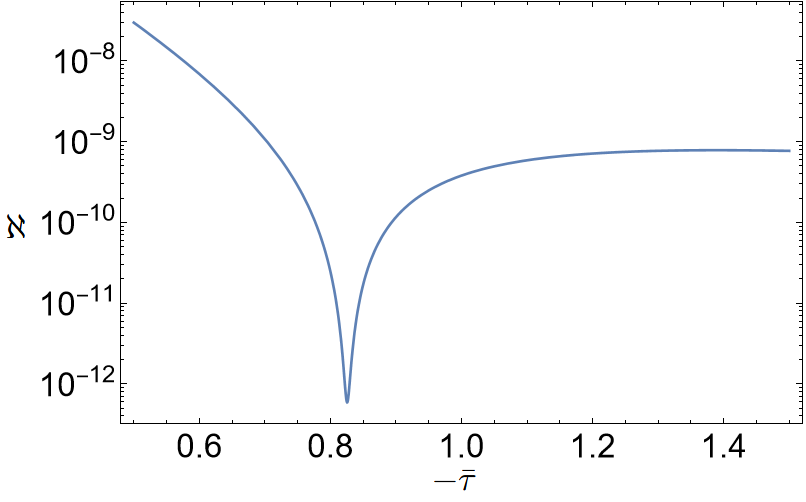}}
\caption{Depictions of the function $\aleph(\bar{\tau})$ for different values of $\mathfrak{m}$ and $\delta\tau$; all considered cases lead to the same qualitative behaviours as shown here. (a): $\mathfrak{m} =\delta\tau =1$, (b): $\mathfrak{m} = 10^4$, $\delta\tau =1$}
    \label{fig:Plots}
\end{figure}

As long as $\aleph > 0$, we know that the right-hand side of Eq.~(\ref{eq:LogArg}) is positive. We can easily see that, in both asymptotic limits $\bar{\tau} \to -\infty$ and $\delta\tau \to 0$, we have $\aleph \to 0$. In addition, since $\tau\geq \tau_0$, the maximum value of $\bar{\tau}$ is $-\delta\tau/2$; $\aleph$ diverges at this value. We have tested the function $\aleph(\bar{\tau})$ for benchmark values of $\mathfrak{m}$ and $\delta\tau$:\ $(\mathfrak{m} = \delta\tau =1)$, $(\mathfrak{m} =1 \gg \delta\tau)$, $(\delta\tau \gg 1 = \mathfrak{m}) $, $ (\delta\tau=1 \gg \mathfrak{m})$, $( 1 \gg \mathfrak{m}, \delta\tau)$, $(\mathfrak{m} \gg 1 \gg  \delta\tau)$, $( \delta\tau\gg 1 \gg \mathfrak{m} )$, $(\mathfrak{m}\gg 1 = \delta\tau)$, and $( \mathfrak{m}, \delta\tau \gg 1)$. For most cases, we find strictly increasing functions of the same qualitative behaviour shown in Fig.~\ref{fig:Plots}(a), while for the last two we have found oscillatory but growing functions; see Fig.~\ref{fig:Plots}(b). In all the cases considered, the function $\aleph(\bar{\tau})$ was found to be positive, suggesting that the right-hand side of Eq.~(\ref{eq:LogArg}) is similarly positive. Consequently, we have the scenario in which the overlap between the vacuum states at different times is zero, such that the Fock bases are unitarily inequivalent, and we do not have a proper unitary evolution from one time slice to the next.

However, this still does not allow us to finally draw a conclusion on whether the VPA indeed vanishes or not. In fact, for this purpose, we also need to demonstrate that the integral in Eq.~(\ref{eq:Q int}), i.e., in the factor $Q(\tau,\tau_0)$, converges. If it instead diverges, it might lead to a non-zero VPA even in the infinite volume limit.

\begin{figure} [htbp]
\centering
   \includegraphics[scale=0.5]{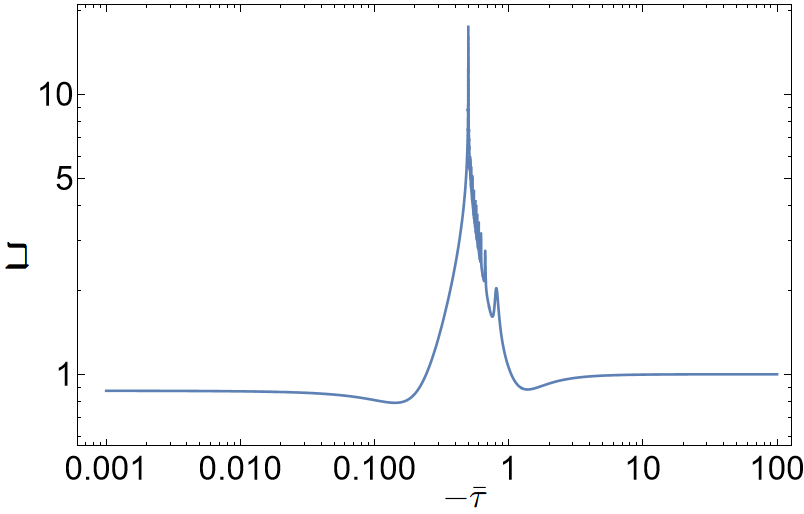}
\caption{Depiction of the function $\beth(\bar{\tau})$ for $\mathfrak{m} =\delta\tau =1$; changing the values of $\mathfrak{m}$ or $\delta\tau$ appears to only shift the figure along the abscissa but does not alter its qualitative features}
    \label{fig:PlotQ}
\end{figure}

The exponent in Eq.~\eqref{eq:Q int} is not Hermitian due to the dependence on $\lambda^+_{\mathbf{k}}(\tau,\tau_0)$. However, the integral will be convergent if the Hermitian part of the matrix $\mathbb{M}_{\mathbf{k}}$, given by
\begin{equation}
\mathbb{M}_{\mathbf{k}}^{\rm H}\coloneqq\frac{1}{2}\left(\mathbb{M}_{\mathbf{k}}+\mathbb{M}_{\mathbf{k}}^{\dag}\right)=\frac{1}{2}\begin{pmatrix} 1+\mathrm{ln}(c_\mathbf{k}(\tau)) && \lambda^{+}_\mathbf{k}(\tau,\tau_0) - \frac{1}{2i}t_\mathbf{k}(\tau) \\ \lambda^{+*}_\mathbf{k}(\tau,\tau_0)+\frac{1}{2i}t_\mathbf{k}(\tau) && 1+\mathrm{ln}(c_\mathbf{k}(\tau))\end{pmatrix}\;,
\end{equation}
is positive definite. This will be the case if the eigenvalues of $\mathbb{M}_{\mathbf{k}}^{\rm H}$ are positive definite. The latter are found to be
\begin{equation}
    m_{\pm}^{\rm H}=\frac{1}{2}\left(1+\ln c_{\mathbf{k}}(\tau)\pm\left|\lambda^{+}_\mathbf{k}(\tau,\tau_0) -\tfrac{1}{2i}t_\mathbf{k}(\tau)\right|\right)\;,
\end{equation}
in which case we require
\begin{equation}
    1+\ln c_{\mathbf{k}}(\tau)>\left|\lambda^{+}_\mathbf{k}(\tau,\tau_0) -\tfrac{1}{2i}t_\mathbf{k}(\tau)\right|\;.
\end{equation}
Analogous to Eq.~(\ref{eq:aleph}), we define a function
\begin{equation}
    \beth :=  1+\ln c_{\mathbf{k}}(\tau) - \left|\lambda^{+}_\mathbf{k}(\tau,\tau_0) -\tfrac{1}{2i}t_\mathbf{k}(\tau)\right|\;.
\end{equation}
As long as $\beth >0$, we know that the eigenvalues of $\mathbb{M}_{\mathbf{k}}^{\rm H}$ are positive definite. In the asymptotic limits, $\bar{\tau} \to -\infty$ and $\delta\tau \to 0$, we obtain $\beth \to 1$. Furthermore, Fig.~\ref{fig:PlotQ} shows that, for $\mathfrak{m}=\delta\tau =1$, this function never reaches values less than or equal to $0$. Using different values for  $\mathfrak{m}$ or $\delta\tau$ only shifts the function along the abscissa. Therefore, we can conclude that $\beth >0$ for all possible parameter values, which in turn implies that the eigenvalues of $\mathbb{M}_{\mathbf{k}}^{\rm H}$ are positive definite and the integral in Eq.~\eqref{eq:Q int} converges. The factor $Q(\tau,\tau_0)$ is therefore finite. Thus, we can then conclude that the VPA is indeed vanishing in the infinite-volume limit for any two values of the conformal time, even if they differ only by an infinitesimal time step.  


\section{Conclusion}
\label{sec:conclusion}

QFT in curved spacetimes suffers from a variety of issues, including the unitary inequivalence of vacua at different times \cite{Giddings:2025abo}. Due to its relevance for applications in cosmology, QFT in de Sitter space constitutes an important testing ground for such problems.

In this article, we have performed an explicit finite-time calculation to illustrate how the unitary time-evolution operator in de Sitter space QFT maps between unitarily inequivalent representations of the canonical algebra. For this purpose, we have considered a free Klein-Gordon scalar field in de Sitter space, performed a canonical quantization, introduced a Fock space, and calculated the vacuum persistence amplitude (VPA). We have shown that the overlap of two vacuum states at different conformal times (the VPA) is zero in the infinite-volume limit, even if there is only an infinitesimal time step between them. Consequently, vacua at different conformal times in de Sitter QFT are unitarily inequivalent. This result constitutes an intensification of the situation in Ref.~\cite{Anderson:2017hts}, where a unitary inequivalence was only predicted to appear in the asymptotic time limit. Where this qualitative difference in both results stems from can easily be seen when realising that Ref.~\cite{Anderson:2017hts} only considers the background dynamics, corresponding to our operator $\hat{S}_S^\dagger(\tau,\tau_0)$, while we also include the Hamiltonian dynamics represented by $\hat{U}_0(\tau,\tau_0)$ via $\hat{S}_S^\dagger(\tau,\tau_0)\hat{U}_0(\tau,\tau_0)$ in our computation. If we ignored the Hamiltonian dynamics or considered asymptotic states such that we could hide the resulting infinite phase factor in the wave function renormalisation, then we would recover the result of Ref.~\cite{Anderson:2017hts}. In addition, our result is confirmed by the general statements made in Ref.~\cite{Giddings:2025abo} and is of similar nature as the one in Refs.~\cite{Blasone:1995zc,Blasone:2025atj}, where a unitary inequivalence in the infinite volume limit has been found in the case of neutrino mixing in flat space. 

We leave considerations of the practical consequences of the lack of a proper unitary evolution for future works. In addition, while we have found compelling indications for the positive definiteness of the real part of the integral multiplying the volume factor in the VPA, an analytical proof for finite times remains elusive, and this may be the focus of future work.


\begin{acknowledgments}
The work of WTE was supported by the University of Manchester. This research was funded in whole or in part by the Austrian Science Fund (FWF) [10.55776/PAT8564023]. The work of PM was supported by a Research Leadership Award from the Leverhulme Trust [Grant No.~RL-2016-028]; a Nottingham Research Fellowship from the University of Nottingham; and a United Kingdom Research and Innovation (UKRI) Future Leaders Fellowship [Grant Nos.~MR/V021974/1 and MR/V021974/2]. For open access purposes, the authors have applied a CC BY public copyright license to any author accepted manuscript version arising from this submission.
\end{acknowledgments}


\appendix


\section{Evolution operator factorisation}
\label{sec:factorisation derivation}

In this appendix, we detail the derivation of the factorisation of the background and free-field evolution operators $S_S(\tau,\tau_0)$ and $U_0(\tau,\tau_0)$, respectively (as presented in Eqs.~\eqref{eq:S product identity} and~\eqref{eq:evo op factorisation}). As these are both built out of creation and annihilation operators, we can treat both within a unified description, specifying the choice of coefficient functions $f^0$, $f^+$ and $f^-$ to recover each one. Thus, taking inspiration from Ref.~\cite{Mitter:1984gt} and the techniques discussed therein, we assume the following ansatz:
\begin{flalign}\label{eq:factorisation ansatz}
    \hat{\mathcal{O}}(\tau,\tau_0;z) \ =& \ \exp\Big[-i\text{Vol}\int_\mathbf{k}\tilde{\lambda}_\mathbf{k}(\tau,\tau_0;z)\Big]\,\exp\Big[-i\int_\mathbf{k}\lambda^{+}_\mathbf{k}(\tau,\tau_0;z)\,\hat{K}^+_{\mathbf{k},I}(\tau_0)\Big]\nn\\ &\ \times\exp\Big[-i\int_\mathbf{k}\lambda^{0}_\mathbf{k}(\tau,\tau_0;z)\,\hat{K}^0_{\mathbf{k},I}(\tau_0)\Big]\,\exp\Big[-i\int_\mathbf{k}\lambda^{-}_\mathbf{k}(\tau,\tau_0;z)\,\hat{K}^-_{\mathbf{k},I}(\tau_0)\Big] \nn\\[1em] \overset{!}{=}&\ \exp\Big[-iz\int_\mathbf{k}\Big(f^0_\mathbf{k}(\tau,\tau_0)\,\hat{K}^0_{\mathbf{k},I}(\tau_0) + f^-_\mathbf{k}(\tau,\tau_0)\,\hat{K}^-_{\mathbf{k},I}(\tau_0)  + f^+_\mathbf{k}(\tau,\tau_0)\,\hat{K}^+_{\mathbf{k},I}(\tau_0)\Big)\Big] \;,
\end{flalign}
where we have factored out any dependence on $e^{i\,\textrm{Vol}\int_\mathbf{k}\tilde{f}_\mathbf{k}(\tau,\tau_0)}$ (with $\tilde{f}_\mathbf{k}(\tau,\tau_0)$ a $c$-numbered function), and $z$ is a real parameter, such that 
\begin{equation}\label{eq:lambdas bc}
     \hat{\mathcal{O}}(\tau,\tau_0;0) \ = \ \hat{\mathds{1}} \qquad\Longrightarrow\qquad \tilde{\lambda}_\mathbf{k}(\tau,\tau_0;0) \ = \ \lambda^{0}_\mathbf{k}(\tau,\tau_0;0) \ = \ \lambda^{\pm}_\mathbf{k}(\tau,\tau_0;0) \ = \ 0\;.
\end{equation}
The operators $\hat{K}^0_{\mathbf{k},I}$, $\hat{K}^-_{\mathbf{k},I}$ and $\hat{K}^+_{\mathbf{k},I}$ are constructed from products of creation and annihilation operators, and are specified in Eq.~\eqref{eq:K operators def}, satisfying the closed algebra~\eqref{eq:K operator algebra}.

We now take the derivative of both sides of Eq.~\eqref{eq:factorisation ansatz} with respect to $z$\footnote{Here, for brevity, we suppress functional dependence on $\tau$, $\tau_0$ and $z$ when it is clear, and restore it otherwise.}:
\begin{flalign}
     \hat{\mathcal{O}}^{-1}\left(\frac{\partial}{\partial z}\hat{\mathcal{O}}\right) \ =& \ -i\int_\mathbf{k}\Big[\text{Vol}\,\tilde{\lambda}'_\mathbf{k}+(\lambda^+_\mathbf{k})'\,\hat{K}^+_{\mathbf{k},I} \: + \: (\lambda^0_\mathbf{k})'\,e^{\int_\mathbf{k}\lambda^{+}_\mathbf{k}\,\hat{K}^+_{\mathbf{k},I}}\,\hat{K}^0_{\mathbf{k},I}\,e^{-\int_\mathbf{k}\lambda^{+}_\mathbf{k}\,\hat{K}^0_{\mathbf{k},I}} \nn\\[0.5em] &\qquad + \: (\lambda^-_\mathbf{k})'\,e^{\int_\mathbf{k}\lambda^{+}_\mathbf{k}\,\hat{K}^+_{\mathbf{k},I}}\,e^{\int_\mathbf{k}\lambda^{0}_\mathbf{k}\,\hat{K}^0_{\mathbf{k},I}}\,\hat{K}^-_{\mathbf{k},I}\,e^{-\int_\mathbf{k}\lambda^{0}_\mathbf{k}\,\hat{K}^0_{\mathbf{k},I}}\,e^{-\int_\mathbf{k}\lambda^{+}_\mathbf{k}\,\hat{K}^0_{\mathbf{k},I}}\Big] \nn\\[0.5em] =& \ -i\int_\mathbf{k}\Big[\text{Vol}\,\left(\tilde{\lambda}'_\mathbf{k} -2i\lambda^+_\mathbf{k}\, e^{-2i\lambda^0_\mathbf{k}}(\lambda^-_\mathbf{k})'\right) + e^{-2i\lambda^0_\mathbf{k}}(\lambda^-_\mathbf{k})'\,\hat{K}^-_{\mathbf{k},I} \nn\\ &\qquad + \left((\lambda^+_\mathbf{k})' - 2i\lambda^+_\mathbf{k}(\lambda^0_\mathbf{k})' - 4(\lambda^+_\mathbf{k})^2\,e^{-2i\lambda^0_\mathbf{k}}(\lambda^-_\mathbf{k})'\right)\hat{K}^+_{\mathbf{k},I} \nn\\ &\qquad + \left((\lambda^0_\mathbf{k})' - 4i\lambda^+_\mathbf{k}\,e^{-2i\lambda^0_\mathbf{k}}(\lambda^-_\mathbf{k})'\right)\hat{K}^0_{\mathbf{k},I}\Big]\;, \\[1em]  \left(\frac{\partial}{\partial z}\hat{\mathcal{O}}\right)\hat{\mathcal{O}}^{-1} \ =& \ -i\int_\mathbf{k}\Big[\text{Vol}\,\tilde{\lambda}'_\mathbf{k}+(\lambda^-_\mathbf{k})'\,\hat{K}^-_{\mathbf{k},I} \: + \: (\lambda^0_\mathbf{k})'\,e^{-\int_\mathbf{k}\lambda^{-}_\mathbf{k}\,\hat{K}^-_{\mathbf{k},I}}\,\hat{K}^0_{\mathbf{k},I}\,e^{\int_\mathbf{k}\lambda^{-}_\mathbf{k}\,\hat{K}^-_{\mathbf{k},I}} \nn\\[0.5em] &\qquad + \: (\lambda^+_\mathbf{k})'\,e^{-\int_\mathbf{k}\lambda^{-}_\mathbf{k}\,\hat{K}^-_{\mathbf{k},I}}\,e^{-\int_\mathbf{k}\lambda^{0}_\mathbf{k}\,\hat{K}^0_{\mathbf{k},I}}\,\hat{K}^+_{\mathbf{k},I}\,e^{\int_\mathbf{k}\lambda^{0}_\mathbf{k}\,\hat{K}^0_{\mathbf{k},I}}\,e^{\int_\mathbf{k}\lambda^{-}_\mathbf{k}\,\hat{K}^-_{\mathbf{k},I}}\Big] \nn\\[0.5em] =& \ -i\int_\mathbf{k}\Big[\text{Vol}\,\left(\tilde{\lambda}'_\mathbf{k} -2i\lambda^-_\mathbf{k}\, e^{-2i\lambda^0_\mathbf{k}}(\lambda^+_\mathbf{k})'\right) + e^{-2i\lambda^0_\mathbf{k}}(\lambda^+_\mathbf{k})'\,\hat{K}^+_{\mathbf{k},I} \nn\\ &\qquad + \left((\lambda^-_\mathbf{k})' - 2i\lambda^-_\mathbf{k}(\lambda^0_\mathbf{k})' - 4(\lambda^-_\mathbf{k})^2\,e^{-2i\lambda^0_\mathbf{k}}(\lambda^+_\mathbf{k})'\right)\hat{K}^-_{\mathbf{k},I} \nn\\ &\qquad + \left((\lambda^0_\mathbf{k})' - 4i\lambda^-_\mathbf{k}\,e^{-2i\lambda^0_\mathbf{k}}(\lambda^+_\mathbf{k})'\right)\hat{K}^0_{\mathbf{k},I}\Big]\;,
\end{flalign}
where $f^{'}(\tau,\tau_0;z)\coloneqq\frac{\partial }{\partial z}f(\tau,\tau_0;z)$. Herein, we have used the Baker--Campbell--Hausdorff formula
\begin{equation}
    e^{z\hat{A}}\,\hat{B}\, e^{-z\hat{A}} \ = \ \hat{B} + z\,[\hat{A},\hat{B}] + \frac{z^2}{2!}\,[\hat{A},[\hat{A},\hat{B}]] + \cdots \;,
\end{equation}
and the operator algebra~\ref{eq:K operator algebra} to re-express the remaining exponential operator terms. 

For consistency, we require that
\begin{flalign}
    \hat{\mathcal{O}}^{-1}\left(\frac{\partial}{\partial z}\hat{\mathcal{O}}\right) \ =& \ \left(\frac{\partial}{\partial z}\hat{\mathcal{O}}\right)\hat{\mathcal{O}}^{-1} \ = \ -i\int_\mathbf{k}\Big( f^0_\mathbf{k}\,\hat{K}^0_{\mathbf{k},I} \: + \: f^-_\mathbf{k}\,\hat{K}^-_{\mathbf{k},I} + f^+_\mathbf{k}\,\hat{K}^+_{\mathbf{k},I}\Big) \;,
\end{flalign}
from which we can infer the following set of differential equations:
\begin{subequations}
    \begin{flalign}
        \tilde{\lambda}'_\mathbf{k} -2i\lambda^+_\mathbf{k}\, e^{-2i\lambda^0_\mathbf{k}}(\lambda^-_\mathbf{k})' \ = \ \tilde{\lambda}'_\mathbf{k} -2i\lambda^-_\mathbf{k}\, e^{-2i\lambda^0_\mathbf{k}}(\lambda^+_\mathbf{k})' \ =& \ 0 \;, \\[0.5em] (\lambda^-_\mathbf{k})' - 2i\lambda^-_\mathbf{k}(\lambda^0_\mathbf{k})' - 4(\lambda^-_\mathbf{k})^2\,e^{-2i\lambda^0_\mathbf{k}}(\lambda^+_\mathbf{k})' \ = \ e^{-2i\lambda^0_\mathbf{k}}(\lambda^-_\mathbf{k})' \ =& \ f^-_{\mathbf{k}} \;, \\[0.5em] (\lambda^+_\mathbf{k})' - 2i\lambda^+_\mathbf{k}(\lambda^0_\mathbf{k})' - 4(\lambda^+_\mathbf{k})^2\,e^{-2i\lambda^0_\mathbf{k}}(\lambda^-_\mathbf{k})' \ = \ e^{-2i\lambda^0_\mathbf{k}}(\lambda^+_\mathbf{k})' \ =& \ f^+_{\mathbf{k}} \;, \\[0.5em] (\lambda^0_\mathbf{k})' - 4i\lambda^-_\mathbf{k}\,e^{-2i\lambda^0_\mathbf{k}}(\lambda^+_\mathbf{k})' \ = \ (\lambda^0_\mathbf{k})' - 4i\lambda^+_\mathbf{k}\,e^{-2i\lambda^0_\mathbf{k}}(\lambda^-_\mathbf{k})' \ =& \ f^0_{\mathbf{k}}\;.
    \end{flalign}
\end{subequations}
For the background evolution operator $\hat{S}_S$, the situation is very simple, corresponding to $f^0_\mathbf{k}=0$ and $f^{\pm}_\mathbf{k}=\mp i\theta_{\mathbf{k}}$. The resulting differential equations can readily solved using the boundary conditions~\eqref{eq:lambdas bc}:
\begin{subequations}
    \begin{flalign}
        \tilde{\lambda}_\mathbf{k}(\tau,\tau_0;z) \ =& \ \frac{i}{2}\,\textrm{ln}\left(\cosh(z\,\theta_\mathbf{k}(\tau))\right) \;,\\[0.5em] \lambda^{\pm}_\mathbf{k}(\tau,\tau_0;z) \ =& \ \pm\frac{i}{2}\,\tanh(z\,\theta_\mathbf{k}(\tau)) \;,\\[0.5em] \lambda^0_\mathbf{k}(\tau,\tau_0;z) \ =& \ i\,\textrm{ln}\left(\cosh(z\,\theta_\mathbf{k}(\tau))\right) \;,
    \end{flalign}
\end{subequations}
leading precisely to the form given in Eq.~\eqref{eq:S product identity}, upon setting $z =1$.

In the case of the free-field evolution operator $\hat{U}_0$, we have $f^0_\mathbf{k}=g_\mathbf{k}$ and $f^{\pm}_\mathbf{k}=h_{\mathbf{k}}$ [cf.~Eq.~\eqref{eq: f, g and h defs} for their exact forms]. By making the canonical choice
\begin{equation}
    \lambda^{\pm}_\mathbf{k}(\tau,\tau_0;z) \ = \ h_\mathbf{k}(\tau,\tau_0)\,X_\mathbf{k}(\tau,\tau_0;z)\;,
\end{equation}
the system reduces to a set of three differential equations
\begin{subequations}\label{eq:reduced lambda eqs}
    \begin{flalign}
        e^{-2i\lambda^0_\mathbf{k}}X'_{\mathbf{k}} \ =& \ 1 \;,\\[0.5em] \tilde{\lambda}'_\mathbf{k} - 2ih^2_{\mathbf{k}}X_{\mathbf{k}} \ =& \ 0 \;,\\[0.5em] (\lambda^0_\mathbf{k})' - 4ih^2_{\mathbf{k}}X_{\mathbf{k}} \ =& \ g_{\mathbf{k}} \;.  
    \end{flalign}
\end{subequations}
The solutions depend upon whether the discriminant function $\Delta_\mathbf{k}\coloneqq\sqrt{g_\mathbf{k}^2-4h_\mathbf{k}^2}$ is real or imaginary. From Eq.~\eqref{eq: f, g and h defs} (recalling that $\tau_0,\tau\in (-\infty,0)$, with $\tau>\tau_0$, and $\mathbf{k}\in (-\infty,+\infty)$), we have 
\begin{equation}
    \Delta_\mathbf{k}(\tau,\tau_0) \ = \ \frac{|\tau-\tau_0|}{3H(\tau_0\tau)^{3/2}}\sqrt{\tau^2+\tau_0\tau+\tau_0^2}\sqrt{3H^2\mathbf{k}^2\tau_0^2\tau^2+m^2(\tau^2+\tau_0\tau+\tau_0^2)}\;.
\end{equation}
It is clear that $\Delta_\mathbf{k}\in\mathbb{R}$ and furthermore $\Delta_\mathbf{k}\geq 0$. As such, the solutions lie on the real branch. Upon imposing the boundary conditions~\eqref{eq:lambdas bc}, the system~\eqref{eq:reduced lambda eqs} can be solved to give
\begin{subequations}
    \begin{flalign}
        \tilde{\lambda}_\mathbf{k}(\tau,\tau_0;z) \ =& \ -\frac{z}{2}\,g_\mathbf{k}(\tau,\tau_0) + \frac{i}{2}\,\textrm{ln}\left(\cos(z\,\Delta_\mathbf{k}(\tau,\tau_0))-\frac{ig_\mathbf{k}(\tau,\tau_0)}{\Delta_\mathbf{k}(\tau,\tau_0)}\sin(z\,\Delta_\mathbf{k}(\tau,\tau_0))\right) \;,\\[0.5em] X_\mathbf{k}(\tau,\tau_0;z) \ =& \ \frac{\tan(z\,\Delta_\mathbf{k}(\tau,\tau_0))}{\Delta_\mathbf{k}(\tau,\tau_0)-ig_\mathbf{k}(\tau,\tau_0)\tan(z\,\Delta_\mathbf{k}(\tau,\tau_0))} \;,\\[0.5em] \lambda^0_\mathbf{k}(\tau,\tau_0;z) \ =& \ i\,\textrm{ln}\left(\cos(z\,\Delta_\mathbf{k}(\tau,\tau_0))-\frac{ig_\mathbf{k}(\tau,\tau_0)}{\Delta_\mathbf{k}(\tau,\tau_0)}\sin(z\,\Delta_\mathbf{k}(\tau,\tau_0))\right) \;.
    \end{flalign}
\end{subequations}
At $z = 1$, these lead to the solutions given in Eq.~\ref{eq:lambda sols}. The full operator factorisation specified in Eq.~\eqref{eq:evo op factorisation} is then recovered by factoring back in the volume term $e^{-i\,\textrm{Vol}\int_\mathbf{k}f_\mathbf{k}(\tau,\tau_0)}$ with $f_\mathbf{k}$ given by Eq.~\eqref{eq: f, g and h defs}.


\bibliography{Bib}
\bibliographystyle{JHEP}

\end{document}